\documentclass{aa}
\usepackage{graphicx}
\usepackage{times}
\usepackage{psfig}
\usepackage{natbib}
\bibpunct{(}{)}{;}{a}{}{,}
\newcommand{\sqdegr}{\raisebox{0.65ex}{\tiny\fbox{$ $}}\,$^{\circ}$}

\def\hide#1{}

\begin{document}

\title{The evolution of galaxy clustering since z=1 from the Calar
Alto Deep Imaging Survey (CADIS)}
\author{S. Phleps\inst{1} \and K. Meisenheimer\inst{1}}
\offprints{S.Phleps (phleps@mpia-hd.mpg.de)}
\institute{
 Max-Planck-Institut f\"ur Astronomie, K\"onigstuhl 17,
   D-69117 Heidelberg, Germany}

\titlerunning{Evolution of galaxy clustering}
\date{Received 15. 11.  2002 / Accepted ?. ?. 2003}

\abstract{
We present results from an investigation of the clustering evolution
of field galaxies between a redshift of $z\sim1$ and the present
epoch. The current analysis relies on a sample of $\sim3600$
galaxies  from the {\bf C}alar {\bf A}lto {\bf D}eep {\bf I}maging
{\bf S}urvey (CADIS). Its multicolor classification and redshift
determination is reliable up to
$I=23^{\rm mag}$. The redshift distribution extends to $z\sim1.1$, with
formal  errors of $\sigma_z\simeq0.02$. Thus the amplitude
of the three-dimensional correlation function can be estimated by
means of the projected correlation function $w(r_p)$. The validity
of the deprojection was tested on the Las Campanas Redshift Survey
(LCRS), which also serves as a ``local'' measurement. We developed a
new method  to overcome the influence of redshift errors on
$w(r_p)$. We parametrise the evolution of the clustering strength with redshift by
a parameter $q$, the values of which give directly the deviation of the evolution
from the global Hubble flow: $\xi(r_{\mathrm {comoving}}=1 h^{-1}
\mathrm {Mpc})=\xi_0 (1+z)^q$. From a subsample of bright galaxies we
find $q=-3.44\pm0.29$ (for $\Omega_m=1$, $\Omega_\Lambda=0$),
$-2.84\pm0.30$ (for $\Omega_m=0.2$, $\Omega_\Lambda=0$), and
$q=-2.28\pm0.31$ (for $\Omega_m=0.3$, $\Omega_\Lambda=0.7$), that is a significant
growth of the clustering strength between $z=1$ and the present epoch.
From linear theory of dark matter clustering growth one would only expect $q=-2$
for a flat high-density model. Moreover, we establish that the measured clustering
strength depends on galaxy type: galaxies with early type $SED$s
(Hubble type: E0 to
Sbc) are more strongly clustered at redshifts $z\ga
0.2$ than later types. The evolution of the amplitude of the two-point
correlation function for these ``old'' galaxies is much slower
($q=-0.85\pm0.82$ for $\Omega_m=0.3$, $\Omega_\Lambda=0.7$).
Since the evolution of the clustering of bright and early type galaxies
seems to converge to the same value in the local universe, we conclude
that the apparent strong evolution of clustering among all bright
galaxies is dominated by the effect that weakly clustered starburst
galaxies which are common at high redshifts $z\approx 1.0$ have dimmed
considerably since then. Thus the true clustering of massive galaxies
is better followed by the early types. This provides both a natural
explanation for the seemingly conflicting results of previous studies and
accords with the absence of ''faint blue galaxies'' in the local universe.
\keywords{Cosmology: large scale structure -- Galaxies: evolution}
}
\maketitle
\section{Introduction}
At the present epoch galaxies are highly clustered and the universe
seems to be homogeneous only on the very largest scales. The question
of how the structures we see today, and how the galaxies embedded in the dark
matter density field have formed and developed, is still one of the most challenging
ones in the field of cosmology.

First systematic analyses of the distribution of galaxies and clusters did not
occur before galaxy catalogues with large numbers of objects were
drawn up -- the first analyses of the clustering properties of
galaxies were based on the Shane-Wirtanen, the Zwicky catalogue,
and the catalogue of Abell clusters, and the results are outlined in a
number of fundamental papers by Peebles (and co-workers)
\citep{PeeblesI,PeeblesII,PeeblesIII,PeeblesIV,PeeblesV,PeeblesVI}.

Only a few years later the CfA survey was completed
\citep{DavisHuchra82,DavisPeebles83}, and the analysis brought a
distribution to light, which was amazingly inhomogeneous -- filaments,
sheets, walls and large voids emerged, and it became clear that the
{\it local} universe is in fact far from homogeneous.

The local universe has later been studied using the largest catalogues
available today, the Las Campanas Redshift Survey
\citep{LasCampanas}, the Sloan Digital Sky Survey
(\citealp{York00,Stoughton02}), and the 2dFGRS \citep{Colless01}.

It was shown by a number of authors that in the local universe Galaxies are {\it biased
  tracers of  mass}, i.e. older galaxies are much more strongly
clustered than young, starforming galaxies, bright galaxies are more
strongly clustered than faint galaxies
(\citealp{Davis76,Lahav92,Santiago92,Iovino93,Hermit96,Guzzo97,Loveday95,Willmer98,Norberg02,Zehavi02}).
The exploration of the
processes which lead to the different clustering of
galaxies of different Hubble types can help us to understand the
interaction between the pure structure growth of the dissipationless
dark matter component and the development of the baryonic matter into
stars and galaxies. In principle, the evolution of the correlation
function can also place constraints on the cosmological parameters
which determine the geometry and dynamics of our universe.
Detailed observations of the evolution of the clustering strength of different
galaxy types have to be compared to model predictions (large $N$-body
simulations including starformation and feedback,
to disentangle dark matter clustering growth and the evolution of the
bias. A first attempt was made by \citet{Kauffmann99II,Kauffmann99I}, who carried
out a semianalytic simulation of galaxy
formation and clustering in a $\Lambda$CDM cosmology, in which they analysed the
clustering evolution of
galaxies with differing luminosity, color, morphology and
starformation rate.

For the investigation of structure {\it formation} and {\it evolution},
measurements of the clustering strength extending to redshifts of
$z\ga 1$ are required. Until
recently, there have been only few ways to study the large
scale structure of the universe at redshifts $z\ga 1$, see for example
\citet{LeFevre96}, and \citet{Carlberg00}. Not only the
shallow depths of most surveys, but also missing redshift information
or too small number statistics have limited the possibilities of
analysing the data with regard to structure  formation. In general,
two different types of surveys have  to be distinguished -- large
angle surveys, which are limited to relatively bright apparent
magnitudes, and  pencil-beam surveys with small, but very deep fields.
Furthermore, one can distinguish between surveys that contain
only a small number of galaxies, but with very accurate redshift
information (deduced from spectroscopy), and surveys that provide
huge catalogues of galaxies, but without or with rather limited
redshift information. The {\bf C}alar {\bf A}lto {\bf D}eep {\bf
I}maging {\bf S}urvey (CADIS), see \citet{Meise02}, is a deep
pencil-beam survey, the output of which at
present is a catalogue of $\sim6000$ classified objects to
$I\leq23^{\mathrm {mag}}$. Around $4000$ of these  galaxies have reasonably
well determined redshifts inferred by means of multicolour
methods ($\sigma_z\approx 0.02$). This unique data base provides the
possibility to investigate
the evolution of galaxy clustering from a redshift of $z\approx1.1$ to the
present epoch.

Usually, structure is described in terms of $n${\it -point correlation
  functions}, the simplest of which is the {\it two-point correlation
  function}. In practice, computing the three-dimensional real-space
two-point correlation function requires very accurate
distances. Peculiar velocities as well as redshift errors distort the
redshift-space relation, and by making the distribution more
Poisson-like, increase the noise. Different methods have been
developed to overcome these problems. If no
redshifts are available at
all, it is possible to obtain information about the three dimensional
distribution of galaxies by deprojecting the two-dimensional {\it angular
correlation function} $w(\theta)$. If peculiar velocities are not
negligible, or the data suffer from large redshift errors, one can use
the deprojection of the {\it projected correlation function}
\citep{DavisPeebles83} to deduce
the clustering strength of the three-dimensional distribution. This
method is used to derive the results of the present paper.

This paper is structured as follows: The {\bf C}alar {\bf A}lto {\bf
  D}eep {\bf I}maging {\bf S}urvey and the data used for the analysis
are described in Section \ref{CADIS}. An introduction into the
fundamental principles of three- and two-dimensional correlation
functions and the deprojection method used in this paper is given in
Section \ref{theory}. In Section \ref{Cluster} we investigate the
evolution of the galaxy clustering from a redshift of $z\sim1.1$ to
today, in Section \ref{Discussion} the results are discussed.
\section{The Calar Alto Deep Imaging Survey}\label{CADIS}
The {\bf C}alar {\bf A}lto {\bf D}eep {\bf I}maging
{\bf S}urvey was established in 1996 as the extragalactic key project of
the Max-Planck Institut f\"ur Astronomie. It combines a very deep
emission line survey carried out with an imaging
Fabry-Perot interferometer with a deep multicolour survey using three
broad-band optical to NIR filters and up to thirteen medium-band
filters when fully completed. The combination of different observing
strategies facilitates not only the detection of emission line objects but
also the derivation of photometric spectra of all objects in the fields
without performing time consuming slit spectroscopy. Details of the
survey and its calibration will be given in Meisenheimer et al. (in
preparation).

The seven CADIS fields measure $\approx 1/30~\sq\degr$ each and are
located at high Galactic latitudes to avoid dust absorption and
reddening. In all fields the total flux on the IRAS 100\,$\mu$m maps
is less than 2\,MJy/sr which corresponds to $E_{B-V} <0.07$. Therefore
we do not have to apply any colour corrections. As a second selection
criterion the fields should not contain any star brighter than
$\approx 16^{\mathrm {mag}}$ in the CADIS $R$ band. In fact the brightest star
in the four fields under consideration  has an $R$ magnitude of
$15.42^{\mathrm {mag}}$. Furthermore, the fields are chosen such that there
should be at least one field with an altitude of at least $45^\circ$
above the horizon of Calar Alto being observable at any time
throughout the year. Among the CADIS fields three equatorial fields
allow follow-up observations with the VLT.

All observations were performed on Calar Alto, Spain, in
the optical wavelength region with the focal reducers CAFOS (Calar
Alto Faint Object Spectrograph) at the 2.2 m telescope and MOSCA
(Multi Object Spectrograph for Calar Alto) at the 3.5 m telescope.
For the NIR observations the Omega Prime camera was used.

In each filter, a set of 5 to 15 individual exposures was taken. The
images of one set were then bias subtracted, flatfielded and cosmic
corrected, and then coadded to one deep sumframe. This basic data
reduction steps were done with the MIDAS software package in combination with the
data reduction and photometry package MPIAPHOT (developed by  H.-J.~R\"oser and
K.~Meisenheimer).
\subsection{Object detection and classification}
Object search is done on the sumframe of each filter using the source
extractor software {\bf SE}xtractor \citep{Bertin96}. The
filter-specific object lists are then merged into a master list
containing all objects exceeding a minimum $S/N$ ratio in any of the
bands. Photometry is done using the program {\it Evaluate}, which has
been developed by Meisenheimer and R\"oser (see
\citealp{MeisenroeserEval}, and \citealp{Roeser91}). Variations in seeing
between individual exposures are taken into account, in order to get
accurate colours. Because the photometry is performed on individual
frames rather than sumframes, realistic estimates of the
photometric errors can be derived straightforwardly.

The measured counts are translated into physical fluxes outside the
terrestrial atmosphere by using a set of
''tertiary'' spectrophotometric standard stars which were established in the
CADIS fields, and which are calibrated
with secondary standard stars \citep{Oke90,eso} in photometric
nights.

From the physical fluxes, magnitudes and colour indices (an object's
brightness ratio in any two filters, usually given in units of
magnitudes) can be calculated. The CADIS magnitude system is described
in detail in \citet{Wolf01a} and \citet{Fried01}.

With a typical seeing of $1\farcs5$ a morphological star-galaxy
separation becomes unreliable at $R\ga 21$ where
already many galaxies appear compact. Quasars have point-like
appearance, and thus can not be distinguished from stars by
morphology. Therefore a classification scheme was developed, which is
based solely on template spectral energy distributions ($SED$s)
\citep{Wolf01a,Wolf01b}. The classification algorithm basically
compares the observed colours of each object with a colour library of
known objects. This colour library is assembled from observed spectra
by synthetic photometry performed on an accurate representation of the instrumental
characteristics used by CADIS.

The spectral library for galaxies is derived
from the mean averaged spectra of \citet{Kinney96}. From
these, a grid
of 20100 redshifted spectra was formed covering
redshifts from $z=0$ to $z=2$ in steps of $\delta z=0.01$ and 100
different intrinsic $SED$s, from old populations
to starbursts ($SED=1$ corresponds to an E0 galaxy, whereas $SED=100$
is a starburst galaxy).

Using the minimum variance estimator (for details see \citealp{Wolf01a}), each
object is assigned a type
(star -- QSO -- galaxy), a redshift (if it is
not classified as star), and an $SED$.  The formal errors in this
process depend on magnitude and type of the
object. For the faintest galaxies ($I>22$) they are of the order of
$\sigma_z=0.02$, and $\sigma_{SED}=2$, respectively.

Note that we do not apply any morphological star/galaxy separation or
use other criteria. The classification is purely spectrophotometric.

Details about the performance and reliability of the classification
are given in \cite{Wolf01a}, and \cite{Wolf01b}.

Rest-frame luminosities can be estimated directly combining the redshift
information and flux
measurements in the $16$ filters; we do {\it not} apply evolutionary corrections.

Four CADIS fields have been fully analysed so far (for coordinates see
Table \ref{coordtab}). We
identified 4540 galaxies with $I\leq 23$. Out of these, 3626 are located in
the redshift range $0.2\leq z\leq 1.07$, where we have analysed their
clustering properties. The number of galaxies per field is given
in Table \ref{coordtab}, together with the number of bright galaxies
($M_B<-18+5 \log h$).

\begin{figure}[h]
\centerline{\psfig{figure=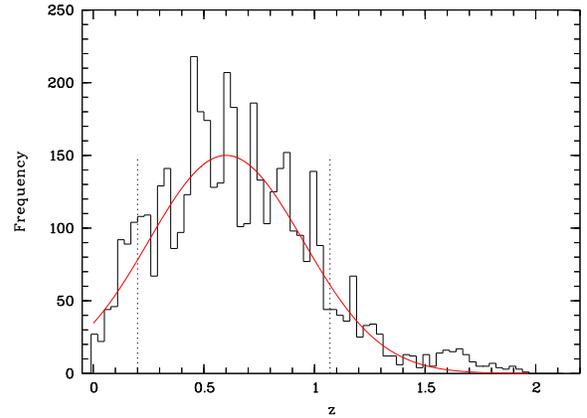,angle=270,clip=t,width=8.3cm}}
\caption[ ]{Redshift distribution galaxies with $I<23$ in the four
CADIS fields. The ''smoothed'' redshift distribution can be described
by a Gaussian. The dotted lines indicate the redshift range, in
which the clustering properties of the galaxies are
analysed.
\label{CADISzdistrib}}
\end{figure}

\begin{table}
\caption[ ]{The coordinates of the four fields investigated so far,
and the number of galaxies per field, $I\leq 23$ and $0.2\leq
z\leq 1.07$, also given is the number of bright galaxies ($M_B<-18+5 \log h$)
in this redshift and apparent magnitude range.\\\label{coordtab}}

\begin{tabular}{r|c c c c }
CADIS field& $\alpha_{2000}$& $\delta_{2000}$& $N_{\mathrm {gal}}$&$N_{\mathrm {gal}}^{\mathrm {bright}}$\\ \hline
1\,h&$1^{\mathrm h} 47^{\mathrm m} 33\fs3$&$2^{\circ} 19' 55''$&$898$&$740$\\
9\,h&$9^{\mathrm h} 13^{\mathrm m} 47\fs5$&$46^{\circ} 14' 20''$&$916$&$727$\\
16\,h&$16^{\mathrm h} 24^{\mathrm m} 32\fs3$ &$55^{\circ}
44' 32''$&$971$&$772$\\
23\,h&$23^{\mathrm h} 15^{\mathrm m} 46\fs9$&$11^{\circ} 27'
00'' $&$841$&$660$\\
\end{tabular}
\end{table}
The redshift distribution (see Figure \ref{CADISzdistrib}) has a
maximum at $z\approx 0.6$. The large variation between adjacent
bins, which is discernably more than expected from
 Poisson statistics,  reflects the large
scale structure of the galaxies. With only four fields, the accumulation of galaxies
at certain redshifts does not cancel out completely. We analyse the
clustering properties of the galaxies between $z=0.2$ and $z=1.07$,
where the statistics are large enough and the redshifts are
reliable.

There has never before existed a sample of this depth and size, that includes
 redshift information.
However, the redshift accuracy is not sufficient to calculate the
three-dimensional two-point correlation function
directly. Nevertheless, as we will demonstrate later, we can use the
redshifts for our analysis. First of all it
enables us to divide the catalogue into distinct redshift bins and we
can analyse the clustering of the galaxies
in each of them up to $z\approx 1.1$. Second, we are able to overcome
the limitations of the angular correlation
function. Instead we can use the {\it projected} correlation function,
which makes {\it direct} use of redshifts,
and is less sensitive to redshift errors than the three-dimensional
 calculation.

We can further divide our sample into different subsamples and
investigate the clustering properties of galaxies
of different $SED$ types, or  different absolute
restframe $B$ magnitudes.
\section{Three- and two-dimensional correlation functions}\label{theory}
The simplest statistic commonly used to describe structure is the {\it
  two-point correlation function} $\xi(r)$.
It describes the excess probability of finding pairs of galaxies
  separated by a distance $r$ over the probability expected in a
  uniform, random distribution of galaxy locations. Thus, the number
  of pairs of galaxies ${\mathrm
d}N_{pair}$ (with one galaxy in the volume element ${\mathrm d}V_1$
  and the other in ${\mathrm d}V_2$) is given in
the form:
\begin{eqnarray}\label{xipair}
{\mathrm d}N_{pair}=N_0^2[1+\xi(\vec{r})]{\mathrm d}V_1~{\mathrm d}V_2~.
\end{eqnarray}
On  intermediate scales the correlation function of galaxies is
featureless. On very small (kpc) scales a change in the slope is
expected \citep{astrophzehavi}, and on scales $20$ to $30$\,Mpc a
cutoff in the angular correlation function has been reported
(\citealp{Maddox90,Collins92,Connolly02}). However, the smooth,
featureless behaviour at the intermediate scales investigated in
the present analysis can be parametrised by a power law, as first
proposed by \citet{Totsuji69}:
\begin{eqnarray}\label{powerlaw}
\xi(r)=\left(\frac{r}{r_0}\right)^{-\gamma}~.
\end{eqnarray}
The {\it correlation length} $r_0$ is a measure for the strength of
the clustering: the probability of finding a
galaxy separated by the distance $r=r_0$ from another galaxy is twice
as large as for a random distribution
of galaxies. Note that the parametrisation by $r_0$ is
equivalent to determining the correlation
amplitude at a specific linear scale.

Since CADIS is a pencil beam survey with a field of view of
$11'\times 11'$, it is of little use to estimate the correlation
length $r_0$ from the measurement, because it is well outside the
measured range of distances. Instead we will deduce the amplitude
of $\xi(r)$ at a comoving separation of $r=1 h^{-1}$\,Mpc.

Throughout this work  correlation functions are determined by using
the estimator invented
by \citet{LandySzalay93}:
\begin{eqnarray}
\xi_{\mathrm {LS}}&=&\frac{\langle DD\rangle-2 \langle RD\rangle+\langle
  RR\rangle}{\langle RR\rangle}~, \label{Landyesti}
\end{eqnarray}
where $\langle DD \rangle$ is the total number of
pairs of galaxies in a given radial bin. $\langle RR \rangle$ is
the histogram of the distances between pairs of randomly distributed
data points that exhibit the same area on the
sky, and suffer from the same selection effects as the real data, thus
acting as a {\it catalogue window}, which
is representing the geometrical properties of the survey. $\langle
RD\rangle$ is the histogram of the distances
between real and random data. Each of the histograms is normalised to
the total number of pairs counted.

Random errors of $\xi$ can be calculated from Poisson statistics \citep{Baugh96}:
\begin{eqnarray}
\sigma_{\mathrm {Poisson}}=\sqrt{\frac{(1+\xi(r))}{[DD]}}~,
\end{eqnarray}
where $[DD]$ is the unnormalised histogram of the number of pairs of
real galaxies in logarithmic distance bins. As \citet{Baugh96} claims,
this may underestimate the errors, but they have found that the
$2 \sigma$ errors computed in this way are comparable to the errors
obtained with bootstrap resampling methods \citep{Barrow84}. As a test
we
calculated bootstrap errors for one cosmological model and one single
redshift bin, and found the
same result. Therefore throughout this paper the errors of all
correlation functions are calculated following
$\sigma_\xi=2\sigma_{Poisson}$.

The real space two-point correlation function can only be calculated
directly, if redshift information is
available with very high precision, and if peculiar velocities are
negligible. This is definitely not the case for
the multicolour data of CADIS. However, it is possible to derive the
parameters by inverting two-dimensional distributions.
\subsection{The angular correlation function}
The angular correlation function $w(\theta)$ is related to the three
dimensional correlation function $\xi(r)$ by
Limber's equation \citep{Limber54}. Limber's equation only holds for
local samples with
$z\ll 1$. For deep samples, which cover a
large range of redshifts, the expansion of the universe, curvature
effects, and the possible evolution of the
clustering have to be included. The redshift-dependent version of
Limber's equation has been derived by
\citet{GrothPeebles77}, and \citet{Phillips78}. The general
distribution of redshifts (${\mathrm d}N/{\mathrm
d}z$) has to be known or calculated from the selection function, which
assembles all selection effects due to
observation and data reduction.

If $\xi(r)$ is assumed to be a power law (equation (\ref{powerlaw})),
the correlation length $r_0$ of the three-dimensional distribution can be
calculated.

The disadvantage of this method is that the inversion is highly
dependent on the redshift selection function
assumed for the survey, which is not a direct observable. Any method
which makes direct use of the
measured redshifts, like the {\it projected
correlation function}, gives much more robust results.
\subsection{The projected correlation function}
If the errors in the redshift measurement are not excessively large,
but still too large to facilitate a direct computation, one can use
the {\it projected correlation function} to derive the correlation
length $r_0$ and the slope $\gamma$ \citep{DavisPeebles83}. It
makes use of the
redshifts, but is much less sensitive to redshift errors than the
three-dimensional correlation function.

For small angles $r^2=r_p^2+\pi^2$.
The projected correlation function is defined as
\begin{eqnarray}\label{projection}
w(r_p)&=& 2\int_0^\infty{\xi\left[(r_p^2+\pi^2)^{1/2}\right]~{\mathrm
d}\pi}\nonumber\\
&=&2\int_{r_p}^\infty{\xi(r)(r^2-r_p^2)^{-1/2}r~{\mathrm d}r}~,
\end{eqnarray}
where $\pi$ is the separation along the line of sight. The projected
physical distance $r_p$ between galaxies $i$ and $j$ (the
distance perpendicular to the line of sight) can -- for a given
cosmology -- be measured very accurately:
\begin{eqnarray}
r_p=[d_A(i)+d_A(j)]\tan(\theta_{ij}/2)~,
\end{eqnarray}
where $d_A$ is the angular diameter distance and $\theta_{ij}$ is the
apparent separation between galaxy i and galaxy j.

If $\xi(r)=(r/r_0)^{-\gamma}$, then equation (\ref{projection}) yields
\begin{eqnarray}\label{wrp}
w(r_p)=C r_0^\gamma r_p^{1-\gamma}~,
\end{eqnarray}
where $C$ is a numerical factor, which depends only on the slope $\gamma$:
\begin{eqnarray}\label{Gamma}
C=\sqrt{\pi}\frac{\Gamma((\gamma-1)/2)}{\Gamma(\gamma/2)}~.
\end{eqnarray}
Thus computing $w(r_p)$ provides a measurement of the parameters of
the three-dimensional correlation function, namely $r_0$ and
$\gamma$. From the amplitude of $w(r_p)$ we can deduce the amplitude of the
three-dimensional correlation function at $r=1 h^{-1}$\,Mpc:

With $\xi(r)=(r/r_0)^{-\gamma}$
\begin{eqnarray}
\xi(r=1 h^{-1}\mathrm {Mpc})=r_0^\gamma
\end{eqnarray}
\begin{eqnarray}\label{ampliatone}
w(r_p)&=&C r_p^{1-\gamma} r_0^\gamma\nonumber\\
&=&C r_p^{1-\gamma}\xi(r=1 h^{-1}\mathrm {Mpc})
\end{eqnarray}
The expressions (\ref{projection}) to (\ref{ampliatone}) refer to
physical distances as would be measured by an observer at the epoch
under consideration. To facilitate a direct comparison of the possible changes with respect
to the global Hubble flow, we compute
the amplitude for a {\it comoving} distance $r_{\rm com}=1 h^{-1}$\,Mpc. With
\begin{eqnarray}
\xi(r_{\rm com}=1 h^{-1}\mathrm {Mpc})&=&(r_0(1+z))^\gamma~,
\end{eqnarray}
which we will call $\xi_{1\mathrm{Mpc}}$ in the following, we find
\begin{eqnarray}\label{supergleichung}
\Rightarrow\xi_{1\mathrm {Mpc}}=\frac{w(r_p)(1+z)^\gamma}{C
r_p^{1-\gamma}}~,
\end{eqnarray}
with the numerical factor C from equation (\ref{Gamma}).\\
The errors are calculated following
Gaussian error propagation:
\begin{eqnarray}\label{gaussianError}
\sigma_{\xi_{1\mathrm {Mpc}}}=\left[\left(\frac{\partial \xi_{1\mathrm {Mpc}} }{\partial w(r_p)}\sigma_{w(r_p)}\right)^2\!\!\!+\!\!\!\left(\frac{\partial \xi_{1\mathrm {Mpc}}}{\partial \gamma}\sigma_\gamma\right)^2\right]^\frac{1}{2}
\end{eqnarray}
In our case the error of $\gamma$ is negligible, as explained in the
following section. The first term completely dominates the error of
$\xi_{1\mathrm{Mpc}}$, thus the approximation $\sigma_{\log\xi_{1
  \mathrm{Mpc}}}=\sigma_{\log A}$ is valid within a fraction of the
statistical error.

 The mean galaxy density is determined from the observed galaxy counts in each
field, which does not necessarily represent the the true density \citep{GrothPeebles77}.
The estimator will be on average biased low with respect to the
true correlation  by a constant amount
\begin{eqnarray}
{\cal I}\approx\frac{1}{S^2}\int{w_{\rm true}(r_p)\mathrm{d}^2
S_1\mathrm{d}^2 S_2}~,
\end{eqnarray}
where $w_{true}(r_p)$ is the true projected correlation function and  $S$
is the physical area corresponding to the solid angle of the field
at the redshift under consideration. If the true correlation function
is given by equation (\ref{ampliatone}), the measurement yields
\begin{eqnarray}
w_m(r_p)&=&C r_0^\gamma  r_p^{1-\gamma}-{\cal I}\label{integconstfit1}\\
&=&C r_0^\gamma  \left[r_p^{1-\gamma}-{\cal I}/(C r_0^\gamma )\right]~.\label{integconstfit2}
\end{eqnarray}
The true amplitude $C r_0^\gamma$ is not known, but
${\cal I}/(C r_0^\gamma )$ can be estimated by performing a Monte Carlo
integration (where we use the mean of the pair counts $\langle
RR\rangle$ at a projected distance $r_p$ of the
four fields):
 \begin{eqnarray}
\frac{{\cal I}}{C r_0^\gamma }=\frac{\sum{\left[\langle RR\rangle\cdot r_p^{1-\gamma}\right]}}{\sum{\langle RR\rangle}}\label{integconstfit3}~.
\end{eqnarray}
The true value of $C r_0^\gamma$ can be estimated by fitting equation
(\ref{integconstfit2}) to the data, taking the value of ${\cal I}/(C
r_0^\gamma)$ from equation (\ref{integconstfit3}). This value,
multiplied by the fitted amplitude $C r_0^\gamma$, yields the integral
constraint ${\cal I}$. This constant will then be added to the measured
amplitude $A$, and from the so corrected amplitude the value of $\xi_{1\mathrm{Mpc}}$
will be inferred by means of equation (\ref{supergleichung}).
\subsubsection{Estimation of the projected correlation function}
Following \citet{DavisPeebles83} one can calculate the projected
correlation function from
\begin{eqnarray}\label{wrpIntegration}
w(r_p)=\int_{-\delta \pi}^{+\delta \pi}{\xi(r_p,\pi)~{\mathrm
d}\pi}~.
\end{eqnarray}
Since the three-dimensional two-point correlation function has the
form of a power law, it converges rapidly to
zero with increasing pair separation. Therefore the integration limits
do not have to be $\pm \infty$, they only
have to be large enough to include all correlated pairs. Since the
{\it direct} observable  is the redshift
$z$ instead of the physical separation along the line of sight, we
make use of a coordinate transformation \citep{LeFevre96}:
\begin{equation}\label{wrpzInteg}
w(r_p)=\int_{-\delta z}^{+\delta z}{\xi(r_p,\pi)\frac{{\mathrm d}\pi}{{\mathrm d}z}~ {\mathrm d}z}~,
\end{equation}
where
\begin{eqnarray}
\lefteqn{\frac{{\mathrm d}\pi}{{\mathrm d}z}=\frac{c}{H_0}\left[(1+z)\left[\Omega_m(1+z)^3\right.\right.}\\\nonumber
&&\left.\left.+(1-\Omega_m-\Omega_\Lambda)(1+z)^2+\Omega_\Lambda\right]^{\frac{1}{2}}\right]^{-1}
\end{eqnarray}
is the increment in {\it physical} distance. We calculate the distance along the
line of sight following \cite{Kayser97}.

In order to determine $w(r_p)$ we use the two-dimensional correlation
amplitude estimated according to \citet{LandySzalay93},
$\zeta_{\mathrm {esti}}(r_p)$ in the following:
\begin{eqnarray}\label{zetaesti}
\zeta_{\mathrm {esti}}(r_p)=\frac{\langle DD(r_p)\rangle_{\delta z}-2\langle RD(r_p)\rangle_{\delta z}+\langle
RR(r_p)\rangle_{\delta z}}{\langle RR(r_p)\rangle_{\delta z}}~,
\end{eqnarray}
where the subscript $\delta z$ implies that projected distances
between pairs of galaxies are only counted if their
separation in redshift space ($\Delta z$) is less than $|\delta z|$.
To derive $w(r_p)$, $\zeta_{\mathrm {esti}}(r_p)$ has to be
multiplied with the ''effective depth'' $\Delta r_\parallel$ in which galaxies
are taken into account:
\begin{eqnarray}\label{zetaIntegration}
w(r_p)&=&\zeta_{\mathrm {esti}}(r_p)\cdot\Delta r_\parallel\nonumber\\
&=&\zeta_{\mathrm {esti}}\cdot\int_{-\delta z}^{+\delta
z}{\frac{{\mathrm d}\pi}{{\mathrm d}z}~ {\mathrm d}z}~.
\end{eqnarray}
In practice, when dealing with large $\delta z$, one has to cope with
a survey selection function of some kind or another between $-\delta z$ and
$+\delta z$, and not with a top-hat function (of probability unity to
find a galaxy within the borders of the survey and zero
otherwise). That is the survey selection function has to be taken into
account when calculating $\Delta r_\parallel$. The
varying probability of finding a second galaxy separated by a redshift
$\Delta z$ from a randomly chosen galaxy can be included in the
calculation by multiplying the integrand in equation
(\ref{zetaIntegration}) with  the smoothed redshift distribution
$N_z^{-1}\mathrm{d}N/\mathrm{d} z$, normalised to unity at
its maximum
($N_z$ is the normalisation constant). Then the integral in equation
(\ref{zetaIntegration}) is modified to
\begin{eqnarray}\label{onedistance}
\Delta r_\parallel=\int_{-\delta z}^{+\delta
z} \frac{{\mathrm d}\pi}{{\mathrm d}z}\left[\frac{1}{N_z}\frac{{\mathrm d}N}{{\mathrm d}z}\right]{\mathrm d}z~.
\end{eqnarray}
With this correction for the selection function, the ``depth''
converges to a fixed value and does not grow infinitely, even if the
integration limits cover more than the total depth of the survey.
When comparing our results with those of other authors who used the projected
correlation function to derive the clustering strength (see for
example \citealp{LeFevre96} or \citealp{Carlberg00}) it has to be kept in
mind that they did not take the the influence of the survey selection
function into account.

The choice of the integration limit $\delta z$ in equation
(\ref{wrpIntegration}) or equation (\ref{onedistance}) is
somewhat arbitrary. To capture the bulk of the correlation signal,
the integration limits should exceed the
redshift separation corresponding to the
correlation length $r_0$, i.e. be larger than the redshift corresponding to
the pairwise velocity dispersion, and of course they have to be larger
than the uncertainty
in the redshift determination.

These requirements set lower limits to the value of $\delta z$. But
how large a $\delta z$ should be chosen? Very large values of $\delta
z$ might more completely integrate the correlation signal, but they do
so at the cost of considerably increased noise, for two reasons:
First, the larger the separation of two galaxies along
the line of sight, the more meaningless (in terms of true distance)
the projected separation perpendicular to the line of sight
becomes. Second, in the extreme case that a pair of galaxies
is separated by physical distances $s\gg r_0$ along the line of
sight, it is most likely not correlated at all, since the correlation
function decreases very fast with distance. Nevertheless, such pairs can display a
very small projected separation and would therefore be regarded as
strongly correlated.

In order to find the appropriate integration limits for our sample, we used the Las
Campanas Redshift Survey, in the following LCRS. The LCRS is
described in detail by \citet{LasCampanas}. The survey
has a median redshift of $\langle z\rangle \sim 0.1$, and
therefore can be regarded as ``local''; the
mean error in redshift is $\sigma_z\approx 2.24\cdot 10^{-4}$, that is
$\sigma_{cz}=67$\,km~s$^{-1}$.

We calculated the projected correlation function of
the six LCRS stripes for increasing integration limits $\delta z$,
fitted the amplitudes at $r_p=0.5 h^{-1}\mathrm {Mpc}$ (the fit was
done in the range $0.07\la r_p\la 2 h^{-1}$\,Mpc ($-1.15\leq \log
r_p\leq 0.3$) to make
sure we fit where the signal to noise is high) and then calculated the
weighted mean of the six stripes. Also the choice of
$r_p=500 h^{-1}$\,kpc allows the most direct comparison with the CADIS data,
where we fit the amplitude at $r_p\approx316 h^{-1}$\,kpc (at the mean
redshift of the survey ($\bar{z}$) this corresponds to a
comoving separation of $\approx505h^{-1}$\,kpc).
Thus, differences in $\gamma$ have little influence on the comparison.

\begin{figure}[h]
\centerline{\psfig{figure=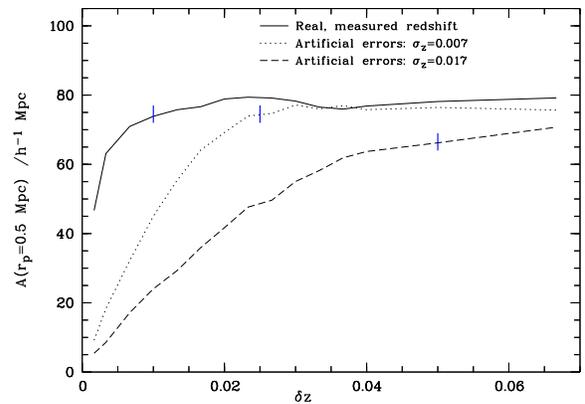,angle=270,clip=t,width=8.3cm}}
\caption[ ]{The influence of redshift measurement errors on the
projected two-point correlation function. $w(r_p)$ for increasing
integration limits is calculated for
 $\sigma_z=0.007$ (dotted line), and
for $\sigma_z=0.017$ (dashed line). The errors of the fits are not
plotted here to avoid confusion.
The  marks indicate the integration limits which have been chosen
for the calculation of $w(r_p)$.
\label{wrpmitfehlern}}
\end{figure}

As can be seen in Figure \ref{wrpmitfehlern}, the estimated amplitude of the projected
correlation function rises very steeply with increasing
integration limits, and approaches a plateau when peculiar velocities become
unimportant and the undistorted correlation signal is sampled.
The maximum value is reached around $\delta z\approx 0.01$, corresponding to
$3000~\mathrm {km s}^{-1}$. Obviously the amplitude of the
projected correlation function reaches its asymptotic value
at the point where the integration limits have about the same size as
the typical velocity dispersion in clusters ($\Delta v\simeq 2.36\cdot
\sigma_v\approx 2500$\,km~s$^{-1}$).

Adding errors in the redshift measurement basically lead to increasing noise
in the correlation signal. If the clustering in redshift space is more
and more washed out (the redshift distribution becomes
more and more Poisson-like), the amplitude decreases,
especially at small scales.

To prove this assumption and to estimate the size of the effect, we
assigned artificial errors to the measured redshifts of the galaxies
in the LCRS catalogue. The errors were randomly drawn out of a
Gaussian error distribution:
\begin{eqnarray}
\hat{z}=z+\Delta z~,
\end{eqnarray}
where $\Delta z$ is randomly drawn from the distribution
\begin{eqnarray}
p(\Delta z)=\frac{1}{\sqrt{2
\pi}~\sigma}\exp\left[-\frac{1}{2}\left(\frac{\Delta z}{\sigma_z}\right)^2\right]~.
\end{eqnarray}
We calculated the estimate of the projected correlation function for the modified
samples, one time with $\sigma_z=0.007$, and one time with
$\sigma_z=0.017$\footnote{representing the core of the distribution of
the CADIS redshift errors as estimated by the comparison of multicolor
redshifts with spectroscopic redshifts for a sample of 160 galaxies}
for increasing integration limits
$\delta z$. For both cases the fitted amplitude of
$w(r_p=500 h^{-1}\mathrm {kpc})$ for increasing integration
limits is also shown in Figure \ref{wrpmitfehlern}.

As expected, redshift errors require larger integration limits for a
stable estimation. For the
artificial redshift errors with $\sigma_z=0.007$, it can be located at
around $\delta z\approx0.025$, for the calculation with
$\sigma_z=0.017$, the maximum is reached at $\delta z\approx0.05$,
that is in both cases at $\delta z\approx 3 \sigma_z$.
From these simulations we conclude that one starts to sample the
correlation signal when the
integration limits are larger than the {\it Full Width at Half Maximum} of
the redshift error distribution ($\Delta z=2\sqrt{2\ln 2}\cdot\sigma_z
\approx 2.3 \sigma_z$). Thus  we conclude
that calculate the projected correlation function
for the CADIS data, the appropriate choice of the integration limits
appears to be $\delta z=\pm0.05$.

Note
that the estimate for the amplitude is systematically lowered in the
presence of significant redshift errors.  For
errors of $\sigma_z=0.017$, we find that the maximum amplitude is a factor 1.2
lower than in the case of the unchanged data.
For a differential comparison between various redshift intervals a
diminished amplitude is in any case no problem, provided one uses the LCRS in its
modified form (with artificial redshift errors simulated exaclty as
those found in the CADIS survey) as local reference.

\section{The evolution of galaxy clustering from $z=1$}\label{Cluster}
\subsection{Clustering properties of all galaxies}
We calculated the projected correlation function for the CADIS data in
three different redshift bins of similar size, $0.2\leq z<0.5$, $0.5\leq z<0.75$,
$0.75\leq z\leq 1.07$.

A random catalogue consisting of 30000 randomly distributed
``galaxies'' was generated  for each CADIS field with the same
properties as the real data. The surroundings of bright stars in our
fields were masked
out. The same mask is applied to the random catalogue.

The calculation was carried out for a
flat high-density model ($\Omega_m=1$, $\Omega_\Lambda=0$), a
hyperbolic low-density model ($\Omega_m=0.2$, $\Omega_\Lambda=0$), and
a flat low-density model with non-zero comological constant ($\Omega_m=0.3$,
$\Omega_\Lambda=0.7$).

\begin{figure}[h]
\centerline{\psfig{figure=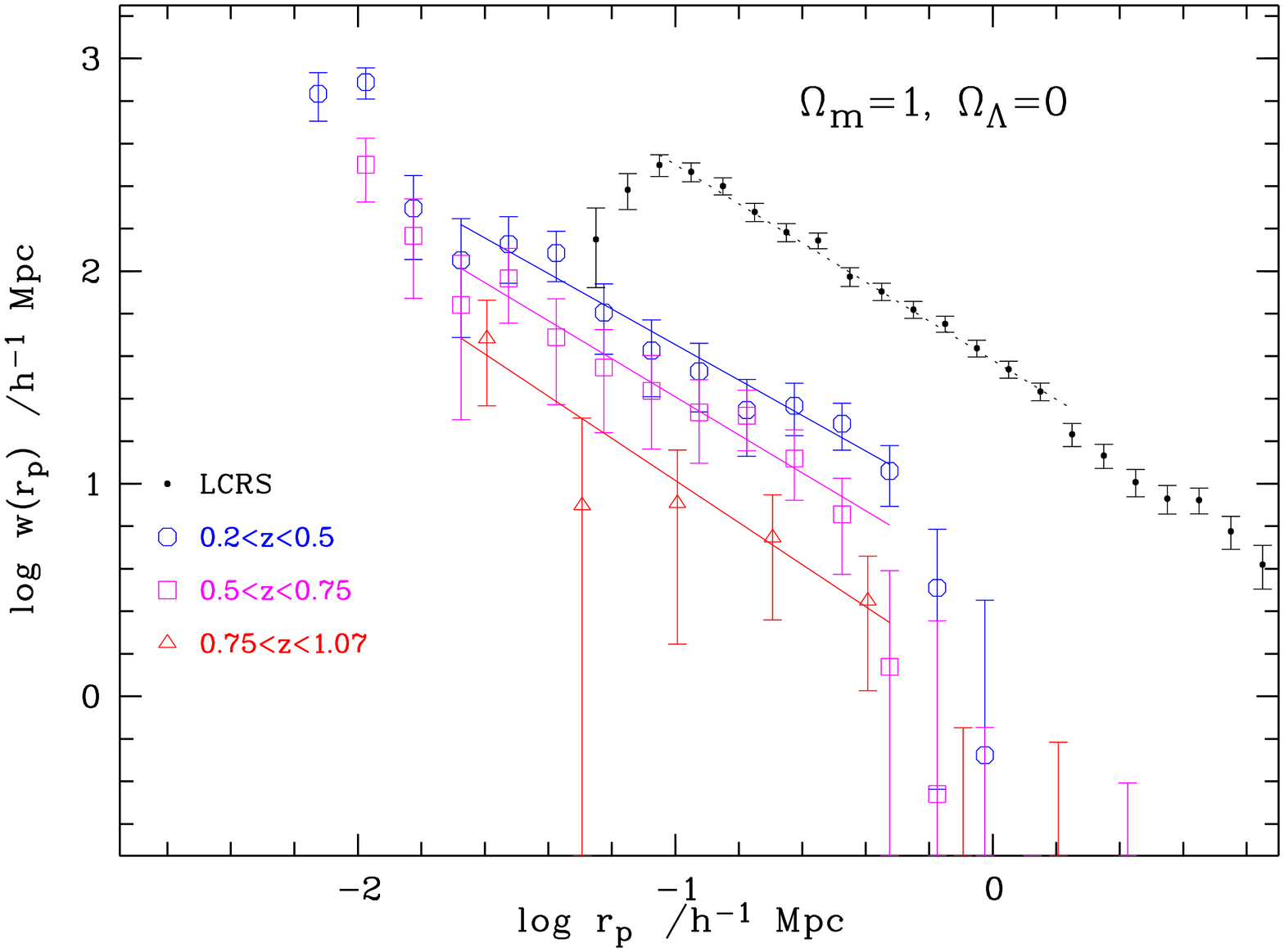,angle=270,clip=t,width=8.3cm}}
\centerline{\psfig{figure=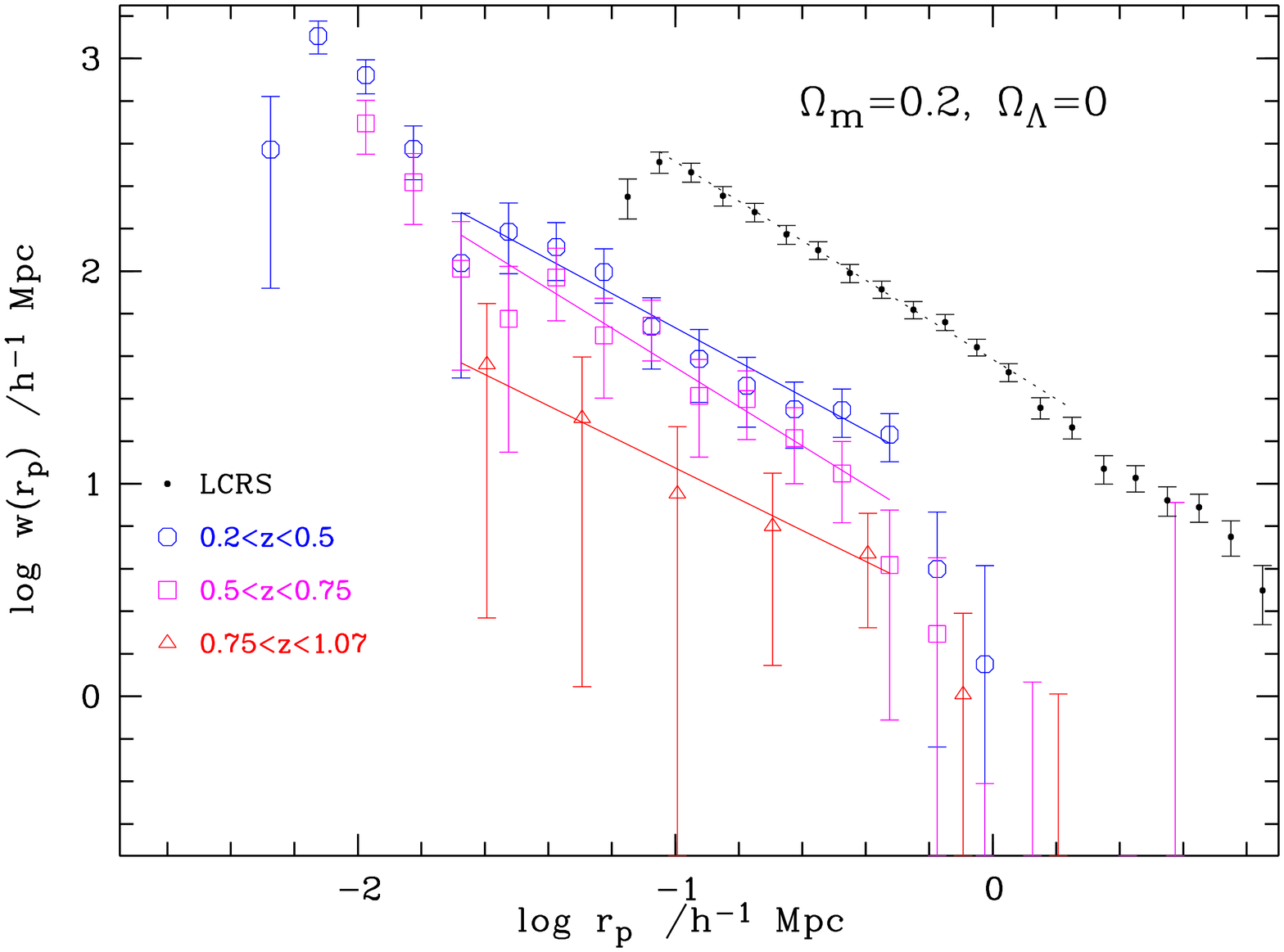,angle=270,clip=t,width=8.3cm}}
\centerline{\psfig{figure=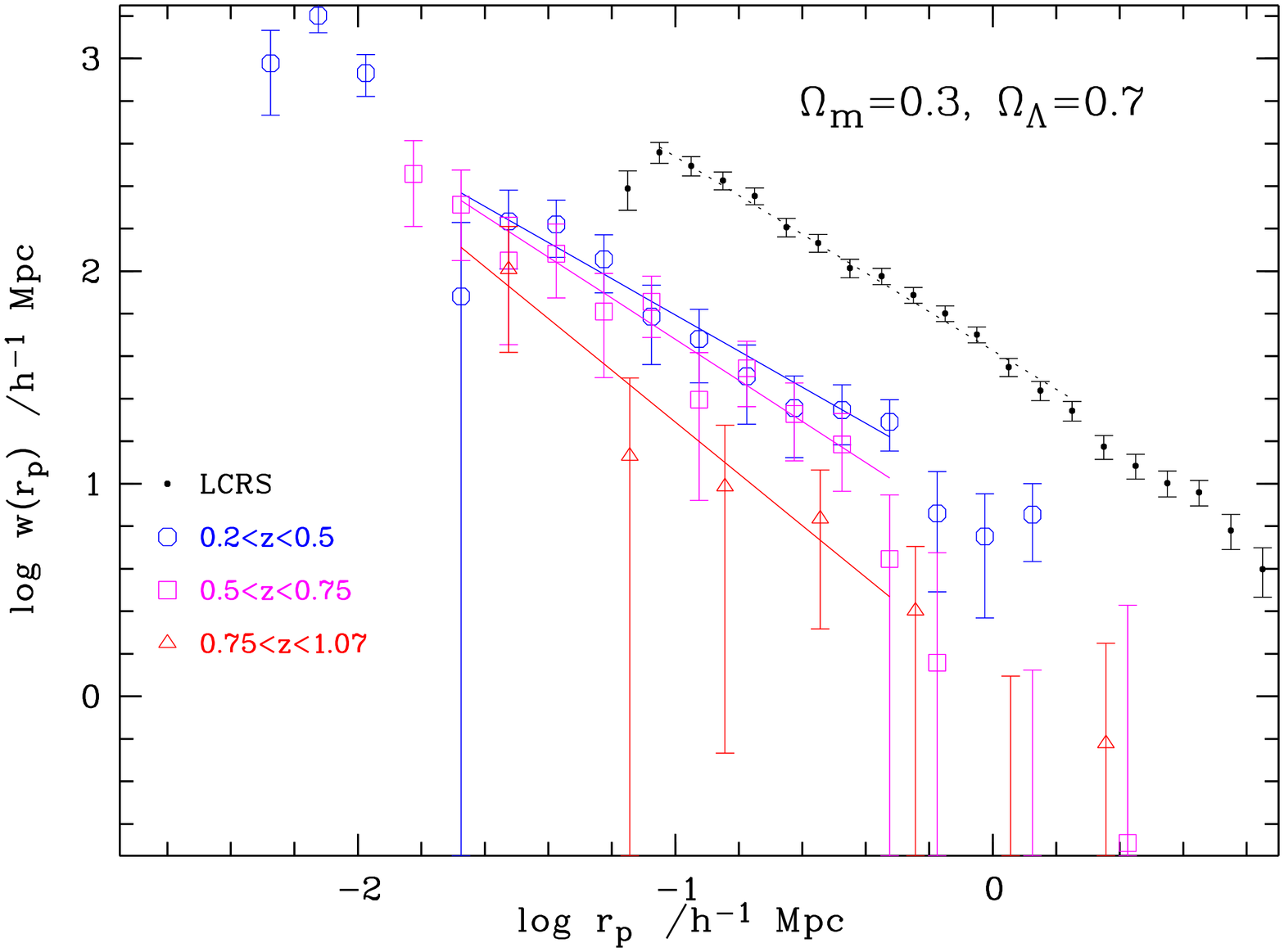,angle=270,clip=t,width=8.3cm}}
\caption[ ]{Measured projected correlation function in three redshift bins, for
the flat standard model (top), an open model (middle), and for
a flat universe with a non-zero cosmological constant (below). Note
that the binning in the highest redshift bin is different.
Note also that the calculation was carried out in proper coordinates as measured
by an observer at the particular redshift under consideration. The
lines are the best fits of the data in the indicated range. \label{projected}}
\end{figure}

\begin{figure}[h]
\centerline{\psfig{figure=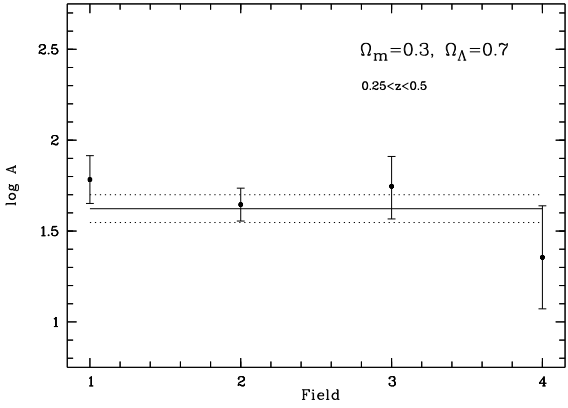,angle=270,clip=t,width=8.3cm}}
\caption[ ]{The amplitudes of the projected
correlation function at $r_p=0.316 h^{-1}$\,Mpc estimated in each individual field, for the bright
galaxies in the lowest redshift bin, for $\Omega_m=0.3$,
$\Omega_\Lambda=0.7$. The solid line is the amplitude measured from
the weighted mean, the dotted lines are the $1 \sigma$.
levels\label{cosmicvariance}}
\end{figure}

For the CADIS data the projected correlation function $w(r_p)$ was
fitted over the range $0.02\la r_p \la 0.5$ $h^{-1}$\,Mpc ($-1.7\leq \log
r_p \leq -0.3$), whereas the LCRS data was fitted in the range
$0.07\la r_p\la 2 h^{-1}$\,Mpc ($-1.15\leq \log r_p\leq 0.3$), see
Figure \ref{projected}. The fitted
amplitudes $A$ at $r_p=316 h^{-1}$ kpc for CADIS and at
$r_p=500 h^{-1}$ kpc for the LCRS, respectively, and the
amplitude of the three-dimensional correlation function at
$r_{\mathrm{com}}=1$\,Mpc derived from them, are listed in Table
\ref{CADISwrptab} (we have left $\gamma$ as a free parameter).\\
Although we correct for the missing variance on large scales by
adding the estimated integral constraint ${\cal I}$ to the measured
amplitudes, cosmic variance may still play a r\^ole. However, as can
be seen from Figure \ref{cosmicvariance}, the errors of the weighted
mean of the correlation function (if at all) only slightly
underestimate the varying clustering strength measured in each single
field. In Figure \ref{cosmicvariance} we show the amplitudes $A$ of the
projected correlation function at $r_p=0.316 h^{-1}$\,Mpc measured in each
individual field (no integral constraints are added yet), for the bright
galaxies in the lowest redshift bin, for $\Omega_m=0.3$,
$\Omega_\Lambda=0.7$.

 The solid line is the amplitude measured from
the weighted mean, the dotted lines are the $1 \sigma$
levels. The spread of the measured values around the mean is
reasonably small, and within their errors they lie within the error of
the weighted mean.

Although the projected distances at which the amplitudes are fitted are
still close to the largest separations measured, the measured amplitude
$A$ and the slope $\gamma$ are independent, the error
$\sigma_\gamma$ is negligible, and therefore the second term in
equation (\ref{gaussianError}) is negligible. Then
\begin{eqnarray}
\sigma_{\log \xi_{1 \mathrm {Mpc}}}=\frac{\partial \log\xi_{1 \mathrm
  {Mpc}}}{\partial \log w(r_p)}\cdot \sigma_{\log w(r_p)}~.
\end{eqnarray}
From equation (\ref{supergleichung}) we find
$\frac{\partial \log\xi_{1 \mathrm {Mpc}}}{\partial \log w(r_p)}=1$,
so
\begin{eqnarray}
\sigma_{\log\xi_{1 \mathrm{Mpc}}}=\sigma_{\log w(r_p)}~.
\end{eqnarray}

The values of the integral constraint, which are then added to the measured
amplitudes of the projected correlation function, are listed in Table \ref{CADISwrptab}.  The
amplitudes of the three-dimensional correlation function at
$r_{\rm com}=1$\,Mpc are derived from the corrected amplitudes of
$w(r_p)$.

\begin{table*}
\begin{center}
\caption[ ]{The measured amplitude of the projected correlation function in
different redshift intervals, at $r_p\approx316 h^{-1}$ kpc for CADIS and at
$r_p\approx500 h^{-1}$ kpc for the LCRS (first lines),
respectively, the integral constraint ${\cal I}$ and $\xi_{1
\mathrm{Mpc}}$ as calculated from the ${\cal I}$-corrected amplitudes
of $w(r_p)$, for different world
models. Numbers in italic indicate that this values are inferred from
the LCRS.\\ \label{CADISwrptab}}
\begin{tabular}{|l|r@{ $z$ }l||r@{$\pm$}l|c|c|c|} \hline
Model &\multicolumn{2}{c||}{$z$ interval}&\multicolumn{2}{c|}{$A$} &
 $\gamma$ &${\cal I}$ & $\xi_{1\mathrm{Mpc}}$\\ \hline\hline
 &{\it 0.04}$\leq$ & $\leq$ {\it 0.16}&{\it
 71.92}&$^{1.83}_{1.79}$&{\it 1.92}$\pm${\it 0.03}&$1.89$&{\it 14.06}$\pm^{0.35}_{0.34}$\\
\cline{2-8}
$\Omega_m=1.0$&$0.2\leq$
 &$<0.5$&$17.27$&$^{2.34}_{2.06}$&$1.83\pm0.09$&$5.53$&$4.23\pm^{0.43}_{0.38}$\\ \cline{2-8}
$\Omega_\Lambda=0$ &$0.5\leq$ &$<0.75$&$9.16$&$^{2.01}_{1.65}$&$1.89\pm0.14$&$3.20$&$3.26\pm^{0.53}_{0.43}$\\ \cline{2-8}
&$0.75\leq$ &$< 1.07$&$3.88$&$^{1.31}_{0.98}$&$2.03\pm0.18$&$1.06$&$1.82\pm^{0.48}_{0.36}$\\ \hline
&{\it 0.04}$\leq$ &$ \leq$ {\it 0.1}&{\it 73.12}&$^{1.91}_{1.87}$&{\it
 1.93}$\pm${\it 0.03}&$1.88$&{\it 14.19}$\pm^{0.36}_{0.35}$\\
\cline{2-8}
$\Omega_m=0.2$&$0.2\leq$
 &$<0.5$&$21.44$&$^{2.79}_{2.47}$&$1.80\pm0.09$&$6.66$&$5.39\pm^{0.52}_{0.45}$\\ \cline{2-8}
$\Omega_\Lambda=0$&$0.5\leq$ &$<0.75$&$12.20$&$^{2.55}_{2.11}$&$1.92\pm0.14$&$3.69$&$4.20\pm^{0.67}_{0.56}$\\ \cline{2-8}
&$0.75\leq$ &$< 1.07$&$5.01$&$^{1.81}_{1.33}$&$1.97\pm0.22$&$1.30$&$2.31\pm^{0.66}_{0.48}$\\ \hline
 &{\it 0.04}$\leq$ & $\leq$ {\it 0.16}&{\it 79.48}&$^{2.00}_{1.95}$&{\it 1.91}$\pm${\it 0.03}&$1.89$&{\it 15.49}$\pm^{0.39}_{0.37}$\\
\cline{2-8}
$\Omega_m=0.3$&$0.2\leq$ &$<0.5$&$23.40$&$^{3.42}_{2.98}$&$1.85\pm0.10$&$6.30$&$5.52\pm^{0.63}_{0.55}$\\ \cline{2-8}
$\Omega_\Lambda=0.7$ &$0.5\leq$ &$<0.75$&$15.72$&$^{3.28}_{2.71}$&$1.97\pm0.13$&$4.30$&$5.31\pm^{0.87}_{0.72}$\\ \cline{2-8}
&$0.75\leq$ &$< 1.07$&$6.19$&$^{2.78}_{1.92}$&$2.11\pm0.25$&$1.33$&$2.81\pm^{1.04}_{0.72}$\\ \hline
\end{tabular}
\end{center}
\end{table*}
\subsection{Clustering evolution of bright galaxies}
\begin{figure}[h]
\centerline{\psfig{figure=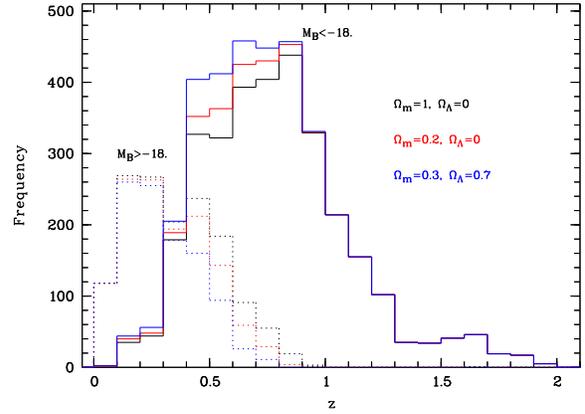,angle=270,clip=t,width=8.3cm}}
\caption[ ]{Redshift distribution of faint ($M_B>-18+5 \log h$)  and bright
($M_B\leq-18+5 \log h$) galaxies, for the different cosmologies under
consideration in this work.\label{cosmozdistrib}}
\end{figure}

It is well known that in the local universe bright
galaxies are much more strongly clustered than faint ones
\citep{Norberg02}, as expected if bright (massive)
galaxies form in the high density peaks of the underlying dark matter
density field \citep{Kaiser84}. For a reliable
determination of clustering {\it evolution} galaxies with similar absolute
luminosities have to be compared. Based on the photometry between
$400$ and $1250$\,nm the absolute
$B$-band luminosities can be determined directly without  any
$K$-correction uncertainty.  At redshifts up to $z\sim 0.3$ CADIS is
dominated by faint ($M_B <-18+5 \log h$) galaxies (see Figure
\ref{cosmozdistrib}), which supress the correlation
signal and distort the measurement of the clustering evolution. A
consistent comparison of galaxy clustering at different redshifts can
be achieved, if we calculate the correlation function in the {\it lowest
two} redshift bins only for galaxies brighter than $M_B=-18+5 \log h$. For this
purpose, the lowest redshift bin has to be confined to $z\ge0.25$
instead of $z=0.2$, because
there are only very few bright galaxies at redshifts smaller than
0.3. At $z=1.07$ the magnitude limit is slightly fainter ($M_B\la -18.6+5 \log h$).

The amplitudes of $w(r_p)$ and $\xi_{1\mathrm{Mpc}}$ are given
in Table \ref{brightones}.

\begin{figure}[h]
\centerline{\psfig{figure=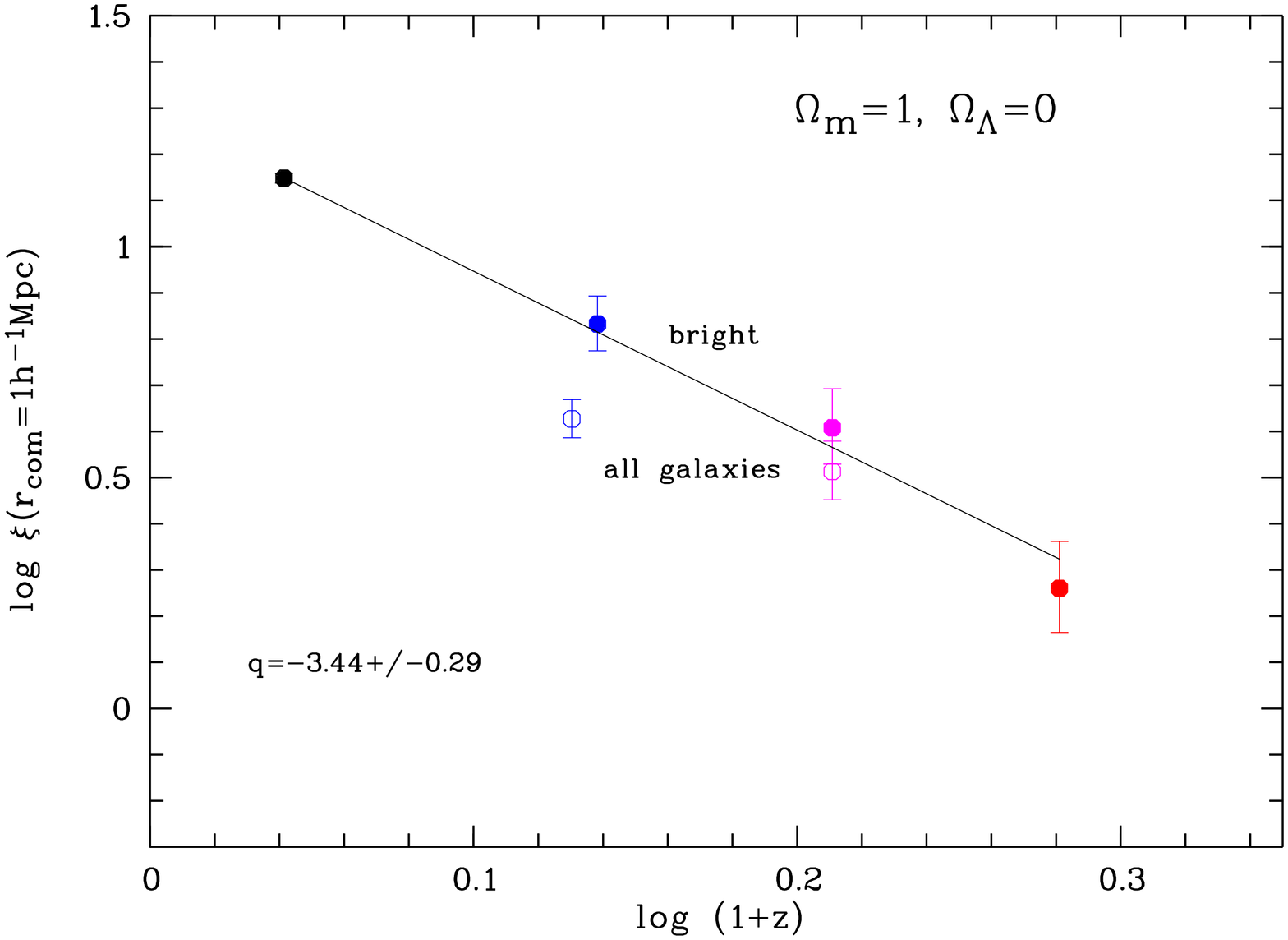,angle=270,clip=t,width=8.3cm}}
\centerline{\psfig{figure=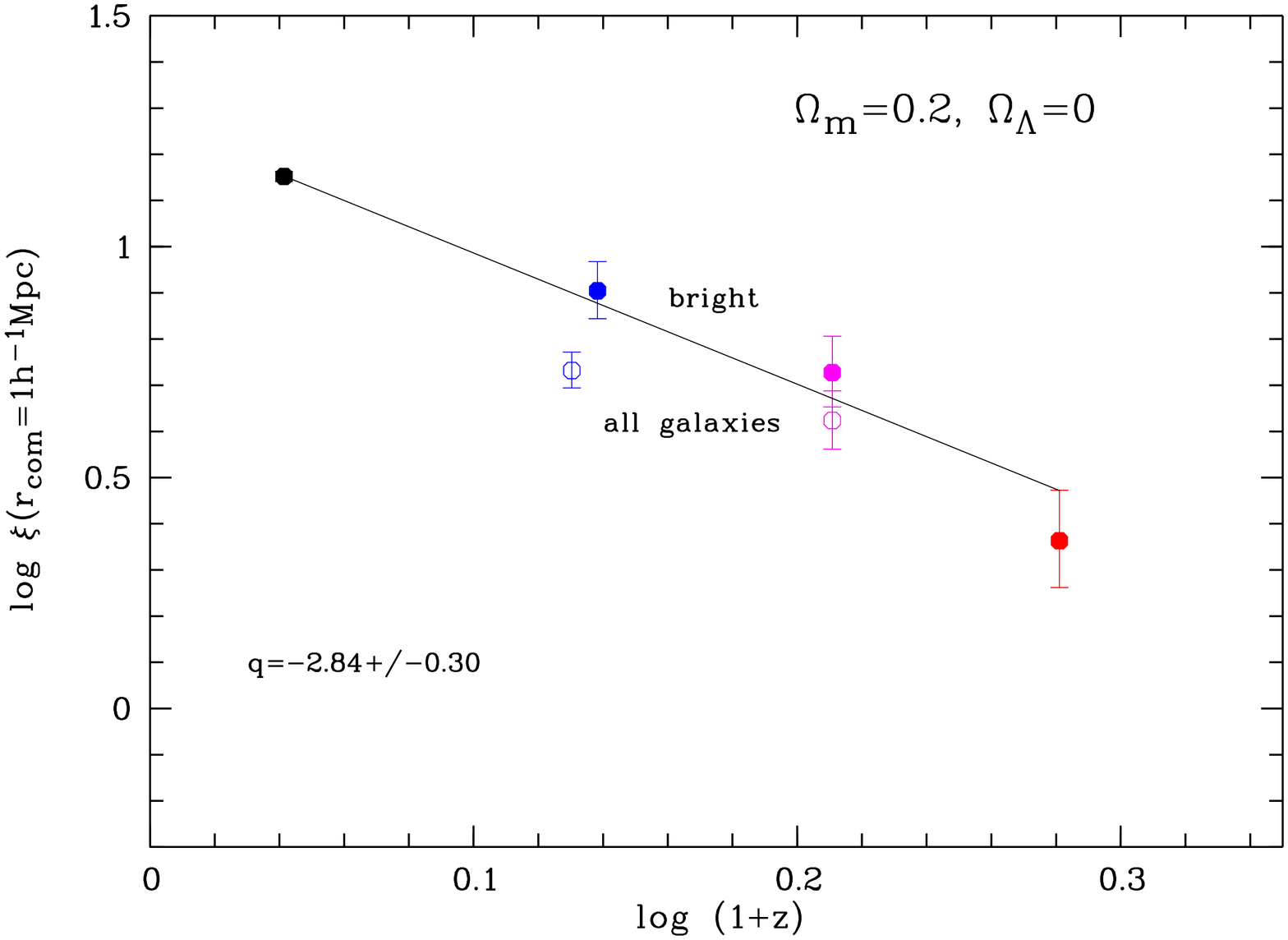,angle=270,clip=t,width=8.3cm}}
\centerline{\psfig{figure=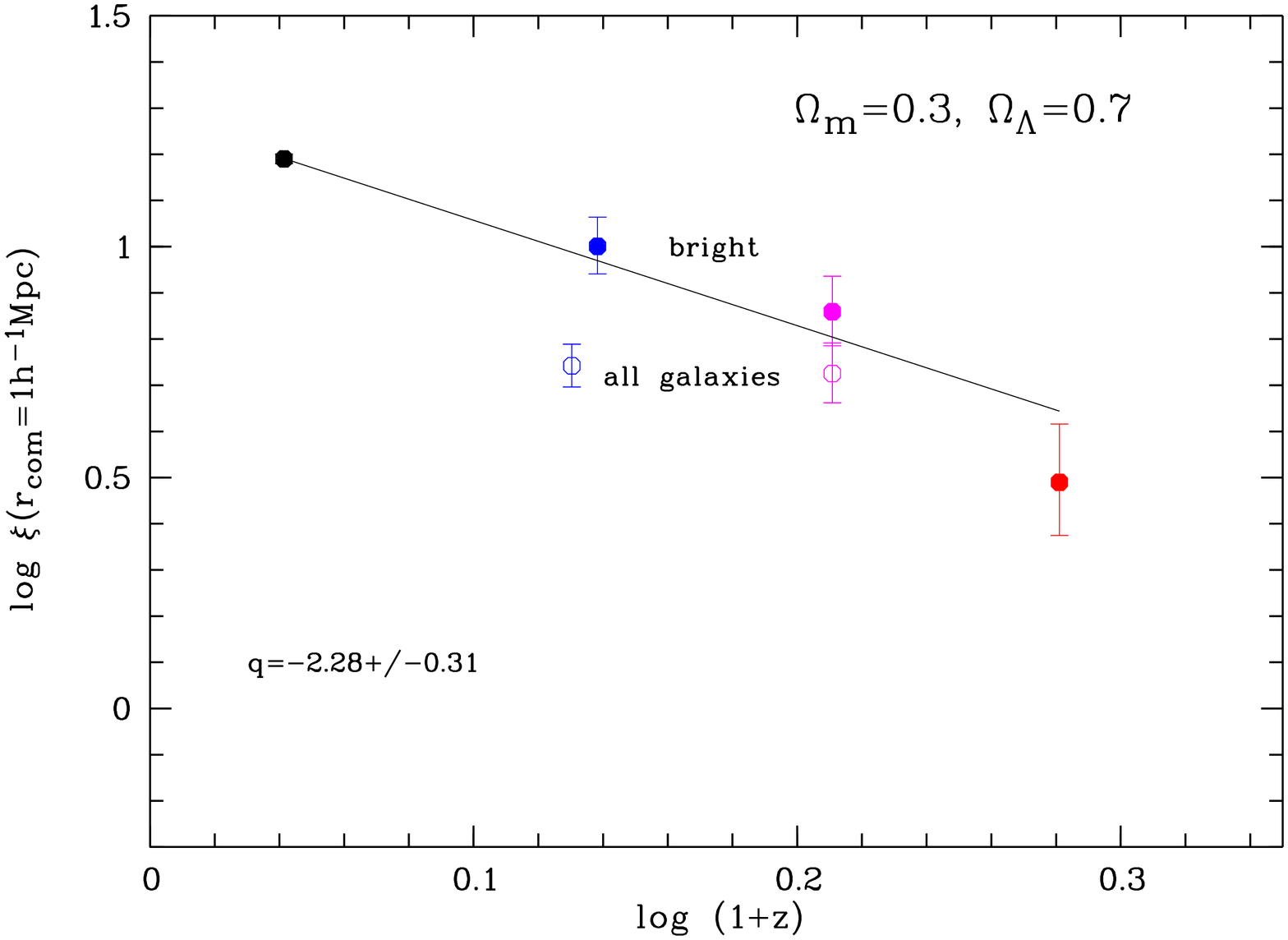,angle=270,clip=t,width=8.3cm}}
\caption[ ]{The evolution of the clustering strength (the amplitude of
  the three-dimensional correlation function at at a {\it
    comoving} separation of $1 h^{-1}$\,Mpc) with
redshift. The integral constraint ${\cal I}$ has been taken into
account. Filled symbols are only bright galaxies, open symbols
include the complete sample. The line is the fit to the bright
galaxies, the first data point (from the LCRS) is included in the
  fit.\label{qparameter}}
\end{figure}

\begin{table*}
\begin{center}
\caption[ ]{The measured amplitude of the projected correlation function at
  $r_p\approx316 h^{-1}$ kpc for the bright ($M_B\leq-18+5 \log h$)
galaxies, the integral constraint ${\cal I}$,
and $\xi_{1\mathrm{Mpc}}$, calculated from the ${\cal I}$-corrected $A$, for different world
models. Note that for the bright
sample the calculation for the first redshift bin was started at
$z=0.25$ instead of $z=0.2$, because there are only very few bright
galaxies at redshifts smaller than 0.3.\\ \label{brightones}}
\begin{tabular}{|l|r@{ $z$ }l||r@{$\pm$}l|c|c|c|} \hline
Model &\multicolumn{2}{c||}{$z$ interval}&\multicolumn{2}{c|}{$A$} &
 $\gamma$ & ${\cal I}$ &  $\xi_{1\mathrm{Mpc}}$   \\ \hline\hline
 &$0.25\leq$ &$<0.5$&$27.38$&$^{5.31}_{4.45}$&$1.98\pm0.14$&$8.26$&$6.80\pm^{1.02}_{0.85}$\\\cline{2-8}
\raisebox{1.5ex}[-1.5ex]{$\Omega_m=1$}&$0.5\leq$&$<0.75$&$11.14$&$^{3.27}_{2.53}$&$2.02\pm0.19$&$4.12$&$4.05\pm^{0.87}_{0.67}$\\\cline{2-8}
\raisebox{1.5ex}[-1.5ex]{$\Omega_\Lambda=0$}&$0.75\leq$&$\leq1.07$&$3.88$&$^{1.31}_{0.98}$&$2.03\pm0.18$&$1.06$&$1.82\pm^{0.48}_{0.36}$\\\hline
&$0.25\leq$ &$<0.5$&$31.50$&$^{6.55}_{5.42}$&$1.85\pm0.17$&$10.24$&$8.02\pm^{1.26}_{1.04}$\\\cline{2-8}
\raisebox{1.5ex}[-1.5ex]{$\Omega_m=0.2$}&$0.5\leq$&$<0.75$&$15.42$&$^{3.98}_{3.16}$&$2.02\pm0.16$&$4.66$&$5.34\pm^{1.06}_{0.84}$\\\cline{2-8}
\raisebox{1.5ex}[-1.5ex]{$\Omega_\Lambda=0$}&$0.75\leq$&$\leq1.07$&$5.01$&$^{1.81}_{1.33}$&$1.97\pm0.22$&$1.30$&$2.31\pm^{0.66}_{0.48}$\\\hline
&$0.25\leq$ &$<0.5$&$39.46$&$^{8.15}_{6.76}$&$1.83\pm0.18$&$12.74$&$10.01\pm^{1.57}_{1.29}$\\\cline{2-8}
\raisebox{1.5ex}[-1.5ex]{$\Omega_m=0.3$}&$0.5\leq$&$<0.75$&$22.21$&$^{5.27}_{4.26}$&$2.00\pm0.16$&$5.85$&$7.23\pm^{1.40}_{1.13}$\\\cline{2-8}
\raisebox{1.5ex}[-1.5ex]{$\Omega_\Lambda=0.7$}&$0.75\leq$&$\leq1.07$&$6.19$&$^{2.78}_{1.92}$&$2.11\pm0.25$&$1.33$&$2.81\pm^{1.04}_{0.72}$\\\hline
\end{tabular}
\end{center}
\end{table*}
We  parameterise the evolution of the clustering strength with redshift
by a parameter $q$, which gives directly the deviation of the
evolution from the global Hubble flow:
\begin{eqnarray}\label{qdef}
\xi_{1\mathrm{Mpc}}=\xi_0(1+z)^q~,
\end{eqnarray}
thus the parameter $q$ can be deduced from a straight line fit in a
$\log\xi$--$\log(1+z)$ plot (see Figure \ref{qparameter}). $q=-2$ is
expected in case of linear clustering growth.
For the flat high-density model we find $q=-3.44\pm0.29$, for the open model we
find $-2.84\pm0.30$, and for the model with non-zero cosmological
constant $q=-2.28\pm0.31$.

The data point of the LCRS is included in the fit for $q$. Its error is
extremely small compared to the CADIS data, and thus it dominates the
fit. If we exclude the LCRS from the fit, we get $q=3.86\pm0.80$,
$q=3.52\pm0.84$, and $q=3.12\pm0.90$ (cosmologies in the same order as
above). The slopes are systematically steeper, although still the same
within their errors. However, the reduced $\chi^2$ of the fits
including the LCRS is smaller than that without.
We regard this as a corroboration
that the connection to the local universe indeed improves the
determination of $q$.

The dependency of the parameter $q$ on the cosmology chosen for the
calculation is as follows: The estimate of
$w(r_p)$ as calculated from equation (\ref{zetaIntegration}) is the
product of $\zeta_{\mathrm {esti}}(r_p)$ (as defined in equation (\ref{zetaesti})), and the
integral given in equation (\ref{onedistance}).
The physical distances $r_p$ are stretched (or compressed, respectively)
corresponding to the cosmology, that means the individual histograms
$\langle DD\rangle$, $\langle RR \rangle$ and $\langle RD\rangle$ are
shifted along the $r_p$ axis. However, the
estimator $\zeta_{\mathrm {esti}}(r_p)$ is unaffected, because all histograms
are shifted in the same way. The integral $\Delta r_\parallel$ of
course changes with cosmology, thus the dependence of the amplitude of
the projected correlation function on cosmology is due to the
integration along the line of sight. Since we fit always at the same
physical scale ($r_p=316 h^{-1}$\,kpc for CADIS), the values of $\xi_{1 \mathrm
  {Mpc}}$ we deduce from $w(r_p)$ show the same dependency on
cosmology. The  differences in $\Delta r_\parallel$ are increasing
with increasing redshift. At our highest redshift bin ($\langle
z\rangle=0.91$) we expect $\frac{\Delta r_\parallel (\Omega_m=0.3,
  \Omega_\Lambda=0.7)}{\Delta r_\parallel (\Omega_m=1,
  \Omega_\Lambda=0)}\approx1.6$, hence we expect (and find) the
amplitudes of the correlation function to differ by the same
factor. The integral constraint only depends on the measured
correlation function (see equations (\ref{integconstfit1}) to
(\ref{integconstfit3}). It is, though estimated with large errors in
the highest redshift bin,  of the order of 25\% of the measured
amplitude at $316 h^{-1}$\,kpc {\it in all cosmologies}, thus this additional
offset does not change the ratio of the amplitudes in different
cosmologies. If we consider the ``local'' measurement not to change with
cosmology, than we expect the parameter $q$ to change (due to the
change in $\xi_{1 \mathrm{Mpc}}$ at the highest redshift bin) by
$\Delta q=\frac{\log 1.6}{\log 1.91 - \log 1.1}=0.93$ between the matter-dominated
flat cosmology and the flat cosmology with non-zero cosmological constant. We find
$\Delta q= 1.16$. The slightly larger difference measured by
us is caused by the bin at $\langle z\rangle =0.625$ (see Figure
\ref{qparameter}.

\subsection{Comparison with other work}
Before analysing the $SED$ type dependent evolution of galaxy
clustering, we compare our results  with
previous attempts to study the clustering evolution
with redshift. In the literature there are essentially only two
investigations of the evolution of
galaxy clustering, that can be compared with the present
work: one analysis by
\citet{LeFevre96} which has been carried out in the
framework of the
{\bf C}anada {\bf F}rance {\bf R}edshift {\bf S}urvey (in the
following CFRS), and one by \citet{Carlberg00}, done
on the CNOC sample
({\bf C}anadian {\bf N}etwork for {\bf O}bservational  {\bf
C}osmology).

\citet{LeFevre96} used the projected correlation function to
investigate the spatial clustering of 591 galaxies
between $0.2\leq z\la 1.1$, in five CFRS fields (for a description of
the survey see \citet{Lilly95} and \citet{Schade95}, respectively) of
approximately the same size as our
CADIS fields. The objects are primarily located in three parallel
strips for each of the five fields, within which almost 100\%
spectroscopic sampling was obtained, separated by regions where few
spectroscopic observations were carried out. The galaxies with
spectroscopic redshifts have $I\leq22.5^{\mathrm {mag}}$. They
computed the projected correlation function
in three redshift bins between $0.2\leq z\leq0.5$, $0.5\leq z\leq0.75$, and
$0.75\leq z\leq 1.0$, with integration limits of $\delta_z=\pm 0.0075$.
For the connection to $z=0$ they took values
of $r_0(z=0)$ from \citet{Loveday95} and \citet{Hudon96}.

\begin{figure}[h]
\centerline{\psfig{figure=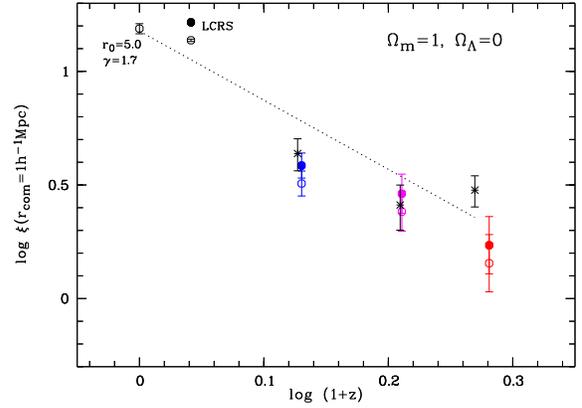,angle=270,clip=t,width=8.3cm}}
\caption[ ]{The amplitudes of the three dimensional correlation
function at $r_{\rm com}=1 h^{-1}$~Mpc, deduced from the projected
correlation function of the CFRS data (crosses), in comparison
with our own data (open symbols). The amplitudes are calculated
without taking the integral constraint into account, to allow for
a direct comparison with  the data of \citet{LeFevre96}. The data
points {\it with} integral constraint correction can be seen in
Figure \ref{qparameter} for comparison. The filled symbols are our
data points corrected for the influence of the redshift errors on
the amplitude of the projected correlation function (shifted by a
factor of 1.2). The dotted line is the fit of the CFRS data points
including the value of $r_0(z=0)$ from \citet{Loveday95}. It seems
that they underestimated the errors, since from a sample of
roughly six times fewer galaxies they derived errors of the same
size as ours.\label{CFRSwrp}}
\end{figure}

Figure \ref{CFRSwrp} shows the amplitude of the three dimensional
correlation function at $r_{\rm com}=1 h^{-1}$~Mpc, deduced from the
projected correlation function, in comparison with our own data.
Since they did not apply corrections for the missing variance, we
calculated $\xi_{1 \mathrm{Mpc}}$ without adding ${\cal I}$ to the
measured amplitudes of $w(r_p)$. For
 direct comparison, we have to multiply our measured amplitudes of
the projected correlation function of the {\it complete sample} by $\frac{A(\delta
  z=0.01)}{A(\delta z=0.05)}=1.2$ in order to allow for the large
 errors of the CADIS multicolour redshifts (see Section \ref{theory}).
With this correction, the CFRS data points are consistent with our own
measurement for the sample as a whole. However, it seems that they
underestimated the errors, since from a sample of roughly six times
less galaxies they derived errors of the same size as ours.

\citet{LeFevre96} claim that if
$r_0(z=0)=5 h^{-1}$~Mpc, $0<\epsilon\la 2$. The fit of their data
points, including a direct connection to $z=0$, that does not take the
different properties of the samples into account, formally yields
$q=-3.04\pm0.21$. If the connection to $z=0$ is disregarded, we
find $q=-1.18\pm0.63$ from their data.

This exercise shows that the value of $q$ derived from the CFRS solely depends on
the connection to the present epoch. To deduce a reliable result, it is
indispensable to take into account that the catalogues to be compared
have to consist of
the same mix of Hubble types and luminosities, as we have seen, bright
galaxies are more  strongly clustered. It is not possible to
estimate the evolution of the clustering strength by comparing a sample of
intrinsically faint galaxies at redshifts $z\ga 0.3$ with the bright
galaxies which dominate the local measurements. The comparison between
different samples has to be carried out in a selfconsistent way.
The integral constraint, which they did not take into account, is
different in different redshift intervals (see Table
\ref{CADISwrptab}), and thus alters the result. Also
the measured amplitude of the
projected correlation function depends on the redshift
accuracy and the appropriate choice of the integration limits.

Thus we conclude that their {\it measurement} is consistent with ours,
but they did not treat the local measurement selfconsistently.

\citet{Carlberg00} carried out an analysis of the clustering evolution
on a sample of 2300 bright galaxies from the CNOC survey. The survey
itself is described in detail in \citet{Yee96}. The galaxies have
$k$-corrected and evolution-compensated $R$ luminosities
$M_R^{k,e}<-20^{\mathrm {mag}}$ ($H_0=100$\,km~s$^{-1}$). The redshift
distribution extends to $z=0.65$. For comparison with $z=0$, they
selected a comparable sample from the LCRS. They also do not
apply corrections for the integral constraint, but since the survey area
of the CNOC is large (extending over $1.55~\sq\degr$ in total)
${\cal I}$ is expected to be negligible at the scales they investigate
($0.16\leq r_p\leq 5.0 h^{-1}$\,Mpc).
The parameter $\epsilon$ (see \citealp{GrothPeebles77,Efstathiou91})
that they use for the parametrisation of the
clustering evolution can of course be related to the parameter $q$
\begin{eqnarray}\label{epsilon}
q=-(3+\epsilon-\gamma)~.
\end{eqnarray}
With equation
(\ref{epsilon}) we can calculate the parameters $q$ which correspond
to the $\epsilon$ values they give: They found $q=-2.00\pm0.22$
for $\Omega_m=1$, $\Omega_\Lambda=0$, $q=-1.37\pm0.18$ for
$\Omega_m=0.2$, $\Omega_\Lambda=0$, and $q=-0.39\pm0.19$ for
$\Omega_m=0.2$, $\Omega_\Lambda=0.8$.

Their values indicate a much slower evolution of the clustering than
our results. The reason for the descrepancy might be the different
sample selection -- the CNOC galaxies have been selected to be bright
in the $R$ band, whereas the galaxies in our sample have bright blue
luminosities. As we will show in the next section, the clustering
growth of blue and red galaxies evolves differently, and in fact the
$q$ values we derived for our subsample of early type galaxies are within
the errors identical  with the values estimated by \citet{Carlberg00}.
\subsection{The evolution of the correlation function for different
  Hubble types}
Various local surveys have established that the red galaxies are
 much more  strongly clustered than the blue ones
(\citealp{Davis76,Lahav92,Santiago92,Iovino93,Hermit96,Guzzo97,Willmer98}).

\citet{LeFevre96} separated their sample into red and blue galaxies
(bluer and redder
than the \citet{Coleman80} Sbc spectral energy distribution), and found that they
have comparable clustering amplitues at $z>0.5$. Here we show that
this is {\it not} the case, at redshifts $z\ga 0.2$ red galaxies are
more
strongly clustered than blue ones.

The galaxy library used for the multicolour classification resembles
regular grids in redshift and {\it SED}, thus the Hubble type can also
be estimated from the observations. This enables us to investigate the
evolution of the clustering of different populations of galaxies.

We devided the sample into two SED bins, with $SED\leq 60$, including
galaxy types E to Sbc (1355 galaxies), and $SED>60$,
including Sds to
Starbursts (2311 galaxies), respectively. Galaxies with earlier $SED$s have smaller
redshift errors, therefore we primarily concentrate on the investigation of
the evolution of the large scale structure of the galaxies in the
$SED\leq60$ sample.

\begin{figure}[h]
\centerline{\psfig{figure=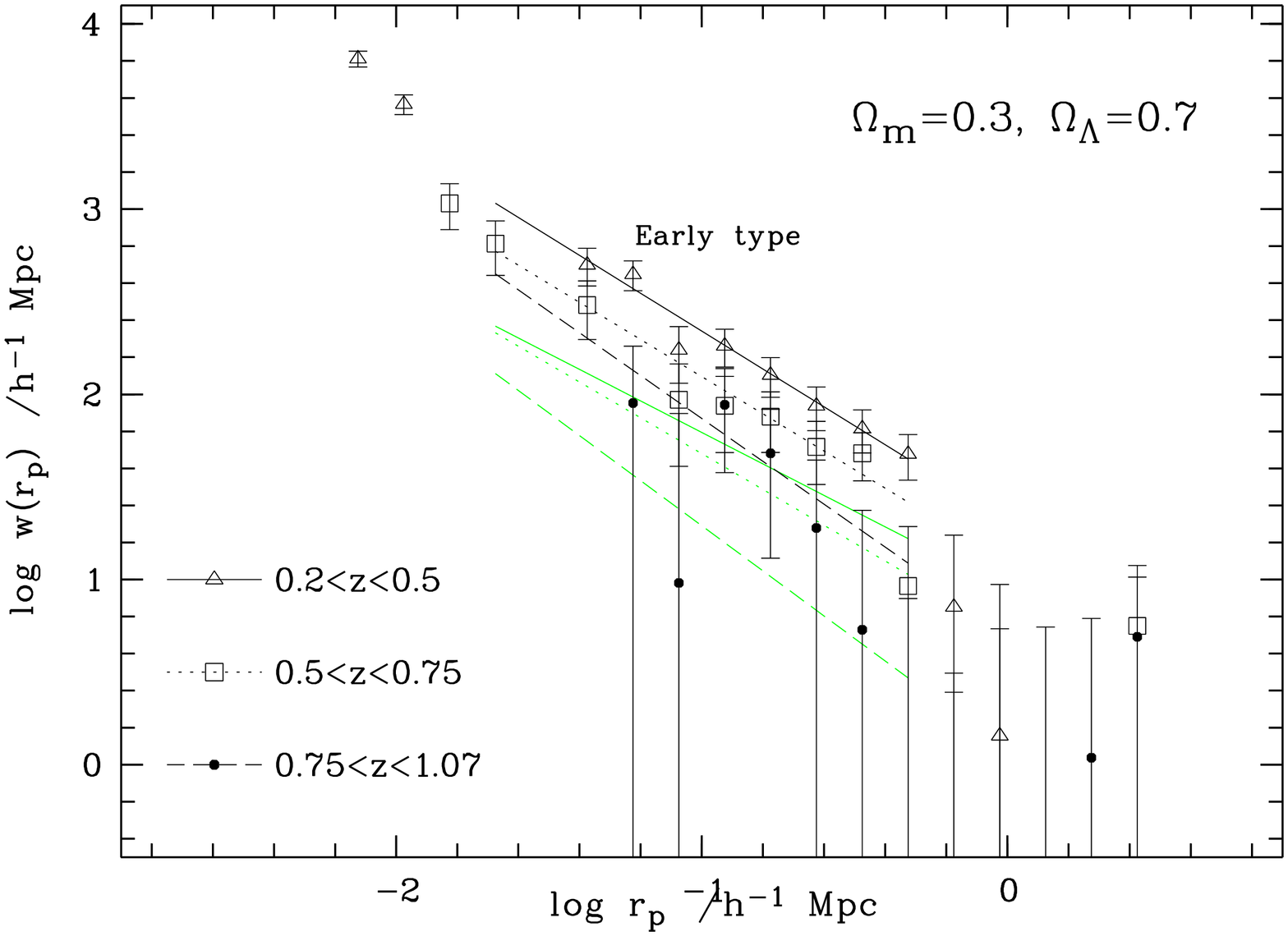,angle=270,clip=t,width=8.3cm}}
\centerline{\psfig{figure=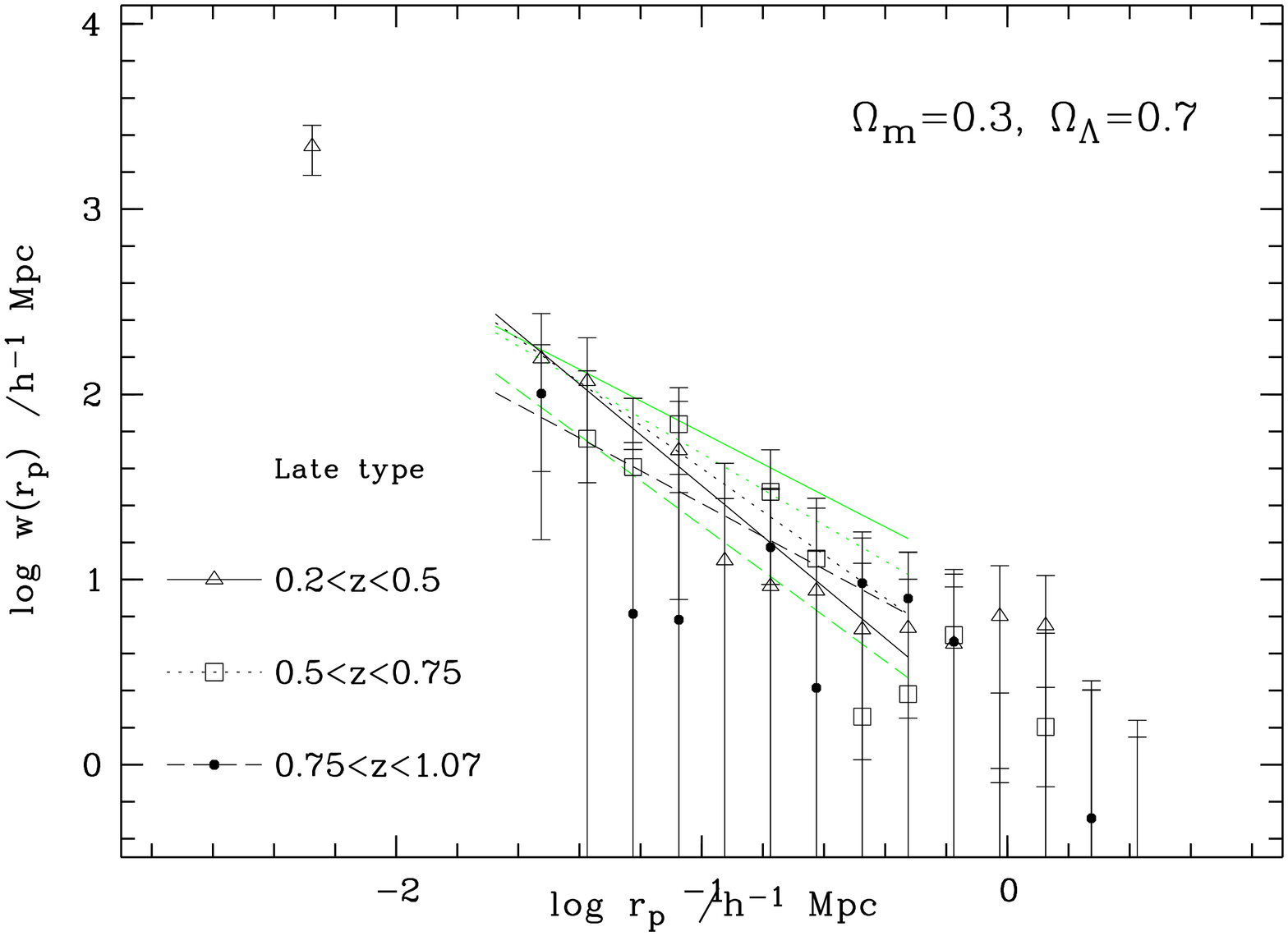,angle=270,clip=t,width=8.3cm}}
\caption[ ]{The projected correlation function in three different
redshift bins; upper panel: early type galaxies in comparison with all
galaxies (only the fits are shown as grey lines), lower panel: late
type galaxies in comparison with all
galaxies for $\Omega_m=0.3$, $\Omega_\Lambda=0.7$.\label{CADISwrpSED}}
\end{figure}

Figure \ref{CADISwrpSED} shows $w(r_p)$ for the different redshift
bins, in comparison with the projected correlation function of all
galaxies, for a flat $\Omega_\Lambda=0.7$
model. The early type galaxies show significantly stronger clustering
than the late types. Table \ref{CADISwrpSEDtab} lists the
amplitudes at $r_p=316 h^{-1}$ kpc, fitted between $\sim 20 h^{-1}$\,kpc
and $\sim 500 h^{-1}$\,kpc, and the integral constraint ${\cal I}$.\\

\begin{table*}
\begin{center}
\caption[ ]{The measured amplitude of the projected correlation function at
$r_p=316 h^{-1}$ kpc and the integral constraint ${\cal I}$, in
different redshift intervals for two SED
bins, for different world models.\\\label{CADISwrpSEDtab}}
\begin{tabular}{|l|l@{ z }r||r@{ $\pm$ }r|c|r||r@{ $\pm$ }r|c|r|} \hline
Model&\multicolumn{2}{c||}{$z$ interval}&\multicolumn{2}{c|}{$A(SED\leq 60)$}&$\gamma$&${\cal I}$&\multicolumn{2}{c|}{$A(SED>60)$}&$\gamma$&${\cal I}$ \\
\hline\hline
&$0.2\leq $&$<0.5$&$46.67$&$^{6.11}_{5.40}$&$1.95\pm0.09$&$14.33$&$7.80$&$^{3.26}_{2,30}$&$1.96\pm0.24$&$0.14$\\ \cline{2-11}
\raisebox{1.5ex}[-1.5ex]{$\Omega_m=1.0$}&$0.5\leq $&$<0.75$&$21.33$&$^{4.40}_{3.64}$&$2.13\pm0.14$&$6.12$&$4.97$&$^{2.94}_{1.85}$&$2.01\pm0.33$&$0.66$\\ \cline{2-11}
\raisebox{1.5ex}[-1.5ex]{$\Omega_\Lambda=0$} &$0.75\leq $&$< 1.07$&$5.53$&$^{9.00}_{3.42}$&$2.84\pm0.93$&$1.01$&$4.78$&$^{1.96}_{1.39}$&$1.98\pm0.22$&$0.57$\\ \hline
&$0.2\leq $&$<0.5$&$59.17$&$^{7.26}_{6.47}$&$1.97\pm0.09$&$15.59$&$9.09$&$^{4.01}_{2.78}$&$1.92\pm0.27$&$2.13$\\ \cline{2-11}
\raisebox{1.5ex}[-1.5ex]{$\Omega_m=0.2$}&$0.5\leq $&$<0.75$&$27.97$&$^{5.39}_{4.52}$&$2.01\pm0.16$&$9.18$&$5.89$&$^{4.35}_{2.50}$&$2.20\pm0.44$&$1.31$\\ \cline{2-11}
\raisebox{1.5ex}[-1.5ex]{$\Omega_\Lambda=0$}&$0.75\leq $&$< 1.07$&$11.03$&$^{10.61}_{5.41}$&$2.69\pm0.80$&$1.42$&$5.67$&$^{2.63}_{1.80}$&$0.52\pm0.86$&--\\ \hline
&$0.2\leq
$&$<0.5$&$67.60$&$^{8.47}_{7.53}$&$2.20\pm0.10$&$15.14$&$6.62$&$^{4.28}_{2.60}$&$2.37\pm0.30$&$1.19$\\\cline{2-11}
\raisebox{1.5ex}[-1.5ex]{$\Omega_m=0.3$}&$0.5\leq $&$<0.75$&$39.22$&$^{7.20}_{6.08}$&$2.00\pm0.11$&$11.12$&$10.40$&$^{7.13}_{4.23}$&$2.16\pm0.44$&$2.30$\\ \cline{2-11}
\raisebox{1.5ex}[-1.5ex]{$\Omega_\Lambda=0.7$ }&$0.75\leq $&$< 1.07$&$19.53$&$^{17.36}_{9.19}$&$2.16\pm0.64$&$3.35$&$9.25$&$^{3.89}_{2.74}$&$1.89\pm0.28$&$1.98$\\ \hline
\end{tabular}
\end{center}
\end{table*}
The corresponding amplitudes of the three-dimensional correlation
function $\xi(r)$ at a comoving distance of $r_{\rm com}=1 h^{-1}$\,Mpc,
corrected for the missing variance ${\cal I}$, are given in
Table \ref{CADISwrpSEDamptab}.

\begin{figure}[h]
\centerline{\psfig{figure=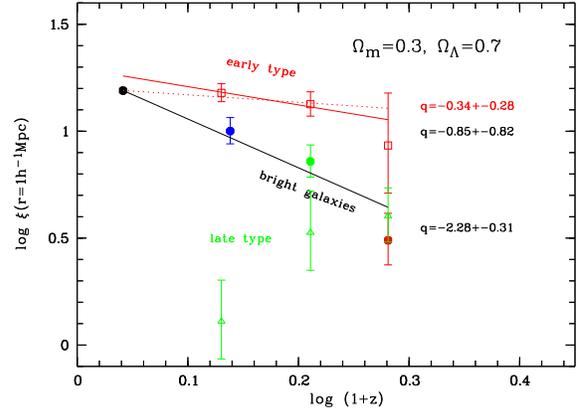,angle=270,clip=t,width=8.3cm}}
\caption[ ]{The evolution of the clustering strength (at a {\it
    comoving} separation of $1 h^{-1}$\,Mpc) with
redshift, for early type ($SED\leq60$). The data
for the bright galaxies and the complete sample are plotted for
comparison. A correction for the integral constraint is included. The
first data point (the weighted mean of the LCRS sectors) is not
included in the fit for the early type galaxies. We also
show the fit which does take it
into account (dotted line), and the measurement for
the late type sample for the $\Omega_m=0.3$,
$\Omega_\Lambda=0.7$ case. Note that the errors of the late type
galaxies are too large to measure the clustering evolution.
\label{CADISwrpSEDampfig}}
\end{figure}

\begin{table*}
\begin{center}
\caption[ ]{The amplitudes of the three-dimensional correlation
function $\xi(r)$ at a comoving distance of $r=1$\,Mpc, for red
($SED\leq60$) and blue ($SED>60$) galaxies; the integral constraint
has been taken into account in the calculation.\\\label{CADISwrpSEDamptab}}
\begin{tabular}{|l|l@{ z }r||c|c|} \hline
Model&\multicolumn{2}{c|}{$z$ interval}&$\xi_{1\mathrm{Mpc}}$
$(SED\leq60)$& $\xi_{1\mathrm{Mpc}}$   $(SED>60)$\\ \hline\hline
&$0.2\leq $&$<0.5$&$11.26\pm^{1.13}_{1.00}$&$1.46\pm^{0.60}_{0.42}$\\ \cline{2-5}
\raisebox{1.5ex}[-1.5ex]{$\Omega_m=1.0$} &$0.5\leq $&$<0.75$&$7.27\pm^{1.17}_{0.96}$&$1.49\pm^{0.78}_{0.49}$\\ \cline{2-5}
\raisebox{1.5ex}[-1.5ex]{$\Omega_\Lambda=0$}&$0.75\leq $&$< 1.07$&$2.34\pm^{3.24}_{1.22}$&$1.96\pm^{0.72}_{0.51}$\\ \hline
&$0.2\leq $&$<0.5$&$13.77\pm^{1.34}_{1.19}$&$2.08\pm^{0.74}_{0.51}$\\ \cline{2-5}
\raisebox{1.5ex}[-1.5ex]{$\Omega_m=0.2$}&$0.5\leq $&$<0.75$&$9.87\pm^{1.43}_{1.20}$&$1.89\pm^{1.15}_{0.66}$\\ \cline{2-5}
\raisebox{1.5ex}[-1.5ex]{$\Omega_\Lambda=0$}&$0.75\leq $&$<
1.07$&$4.56\pm^{3.91}_{1.98}$& --\\ \hline
&$0.2\leq $&$<0.5$&$15.11\pm^{1.56}_{1.37}$& $1.29\pm^{0.72}_{0.43}$\\ \cline{2-5}
\raisebox{1.5ex}[-1.5ex]{$\Omega_m=0.3$}&$0.5\leq $&$<0.75$&$13.37\pm^{1.91}_{1.61}$&$3.35\pm^{1.89}_{1.12}$\\ \cline{2-5}
\raisebox{1.5ex}[-1.5ex]{$\Omega_\Lambda=0.7$} &$0.75\leq $&$< 1.07$&$8.57\pm^{6.51}_{3.44}$&$4.01\pm^{1.42}_{0.98}$\\ \hline
\end{tabular}
\end{center}
\end{table*}
Figure \ref{CADISwrpSEDampfig} shows the amplitudes of the
three-dimensional correlation function $\xi(r)$ at a comoving distance
of $r=1 h^{-1}$\,Mpc, for red galaxies ($SED\leq60$), for
$\Omega_m=0.3$, $\Omega_\Lambda=0.7$. For comparison the
data points for the whole sample are also plotted. The amplitudes are
corrected for the missing variance ${\cal I}$. The lines are the
fits for the $q$-parameter (equation (\ref{qdef})).

The late type galaxies have large redshift errors, and they are
intrinsically only very weakly clustered. Although their
number density is high, the large errors and the low correlation signal
result in a highly
uncertain measurement of their clustering amplitudes (the measurement is shown for
the $\Omega_m=0.3$, $\Omega_\Lambda=0.7$ case), which would make a
reliable estimate of the clustering evolution impossible. Also the late type sample is
at high redshifts dominated by relatively bright galaxies, whereas at
lower redshifts the rather low-mass,
fainter galaxies, which are only very weakly clustered, are currently undergoing
periods of high starformation rates. A galaxy classified as extreme
late type at $z=1$ may have evolved
into an Sa at $z=0$, thus we are not dealing with a homogeneous
sample. A measurement of the clustering evolution of late type
galaxies is therefore not meaningful at this point.

The LCRS measurement is not included in
the fits for the early type galaxies. Table \ref{qparaSEDtab} lists
the parameter $q$ for the early type
sample, and compares them to the values found for the bright
sample. The evolution of the amplitude of the correlation function
is clearly differential, the clustering strength of early type
galaxies grows much slower. However, it looks as if the
clustering evolution of the CADIS early type sample converges to the
same value. In Table \ref{qparaSEDtab} we list also
the parameter $q$ for the early type
sample which was determined when taking the LCRS point into
account.

\begin{table*}
\caption[ ]{The parameter $q$ for early type ($SED\leq60$) galaxies,
  for a fit done without taking the the LCRS point into account, and
  for a fit including it. The evolution of the bright sample
  is shown for comparison. \\\label{qparaSEDtab}}
\begin{center}
\begin{tabular}{l|c c c}
Cosmology& $q(SED\leq60)$, without LCRS&$q(SED\leq60)$, with
LCRS&$q_{\rm bright}$\\\hline
$\Omega_m=1.0$, $\Omega_\Lambda=0.0$&$-2.60\pm0.91$ &$-1.52\pm0.30$ &$-3.44\pm0.29$\\
$\Omega_m=0.2$, $\Omega_\Lambda=0.0$&$-2.03\pm0.82$ &$-0.74\pm0.28$ &$-2.84\pm0.30$\\
$\Omega_m=0.3$, $\Omega_\Lambda=0.7$&$-0.85\pm0.82$ &$-0.34\pm0.28$ &$-2.28\pm0.31$\\
\end{tabular}
\end{center}
\end{table*}
When the LCRS point (at $z=0.1$) is
included in the fit, the CADIS points still lie well on the fitted
line, which indicates that the clustering strength of bright and early
type galaxies indeed seems to converge to the same
point at $z\la0$.

These values of $q$, $-1.47\pm0.37$, $-0.71\pm0.31$, $-0.34\pm0.32$
(see Table \ref{qparaSEDtab} for the corresponding
cosmological models) are within their errors identical to the values
deduced by \citet{Carlberg00} from the red CNOC galaxies (they found
$q=-2.00\pm0.22$, $q=-1.37\pm 0.18$, $q=-0.39\pm 0.19$ for the
cosmologies in the same order as above, see
previous section).

\section{Discussion}\label{Discussion}
We now want to discuss the most important results of our present
analysis in turn:

(1.) Bright galaxies are more strongly clustered than faint galaxies and show
significant growth in clustering since $z \simeq 1$.

At redshifts around $z\approx0.3$ the CADIS data are dominated by
rather faint galaxies, which show smaller clustering amplitudes than
brighter ones. On the other hand, the LCRS sample consists mainly of
intrinsically bright galaxies. In order to be able to compare our high
redshift data self-consistently with the local measurement from the
LCRS, we measured the amplitude of the correlation function from a
subsample of bright galaxies ($M_B\leq -18 + 5 \log h$). We find a
significant growth of the clustering strength for this bright
subsample:

For the $\Omega_m,\Omega_\Lambda = 0.3,0.7$ cosmology we
find $q = -2.2$ that is ${\xi_{\rm 1Mpc} (z=0) \over \xi_{\rm 1Mpc}
(z=1)} = 5.5$. This might be compared with the theoretical predictions
by \citet{Kauffmann99II}.  They use N-body simulations of dark matter
(DM) to derive the clustering of DM halos at $0 < z < 5$ and
semi-analytic models in order to assign a galaxy of certain type and
luminosity to each DM halo. Although many properties of their mock
galaxies population are only in loose accordance with observations,
their result seems robust: the clustering strength of
bright galaxies ($M_B\leq -19 + 5 \log h$) on 2 Mpc scales follows
closely that of the DM halos they are imbedded out to  $z \simeq 1$
(bias parameter $\simeq 1$).
They find a 3.5-fold increase of $\xi_{\rm 2Mpc}$ between
$z=1$ and $z=0$. If one takes into account that the increase should be
higher on 1 Mpc scales and that our limit for ``bright galaxies'' is 1
magnitude fainter ({\it i.e.} it includes less massive halos) it seems
that the
the clustering evolution predicted by $\Lambda$CDM models is in agreement with
our measurements.

(2.) At high redshifts, early type galaxies are more strongly clustered
than the bright galaxies and the clustering strength of the early type
galaxies evolves much more slowly than that of the bright galaxies.

A plausible explanation for
the different evolution of the clustering properties of
the early type galaxies arises in the context
of biased galaxy formation \citep{Bardeen86,Brainerd94}. The first galaxies are born
in a highly clustered state, because they form in the bumps and
wiggles which are superimposed on the very large-scale density
enhancements. The next generations of galaxies
form later in the wings of the large-scale enhancements, and are
therefore less and less
clustered. While the universe expands, the galaxies evolve, age, and eventually
merge to form larger, brighter galaxies and ellipticals, and generally add
to the population of earlier type galaxies, while new galaxies form at
later times in less and less clustered environments.
Merging creates galaxies, which suddenly'' add to the old
population. A merger event also
reduces not only the number of galaxies, but also the number of small
pair separations in a sample, which reduces the probability of finding
pairs of galaxies at small distances -- and thus supresses the
amplitude of the
correlation function. \citet{Fried01} found the density evolution
of the early type and the late type galaxy population in the CADIS galaxy sample
 suggestive of merging.

Although the clustering strength of
the underlying dark matter density field increases with redshift, the
biasing decreases. The net effect is a very slowly rising clustering
amplitude of the early type (old) galaxies.

Thus, the measured rate of the clustering growth depends on
the mixture of galaxy types one observes at different redshifts.

(3.) The evolution of the clustering amplitude of early and bright
galaxies converges to the same value at $z\approx0$.

The convergence of the evolution of the clustering of bright galaxies
and galaxies with $SED$s earlier than Sbc towards the same local
measurement can also be understood if one takes into account that
galaxies evolve and hence the population mix one observes changes with
redshifts.  The comoving number density of weakly clustered starburst
galaxies increases with increasing redshifts,
%
whereas the space density of the
highly clustered very early type (E-Sa) galaxies decreases by a
factor of $\sim1.6$ from $z=0$ to $z=1$ \citep{Fried01}. In our
highest redshift bins the clustering signal of the bright subsample is
presumably dominated by a weakly clustered population of galaxies,
which at that time were {\it bright, and blue}.  At $z=0$ these weakly
clustered blue galaxies have vanished, and now the majority of the
early type galaxies (which had later, bluer $SED$s at higher
redshifts), are also {\it bright}, thus at the present epoch there is
a large overlap between the bright and the early type samples. Thus we
conclude that the apparent strong growth of clustering of the bright
galaxies is dominated by the fading of the unclustered, blue
population.

To understand exactly how galaxy evolution influences
the {\it measurement} of the growth of structure, and the evolution of
the large scale structure influences the evolution of galaxies, we
need detailed investigations of the evolution of both the correlation
function and the luminosity function of galaxies with different
$SED$s.

 Larger, wide angle deep surveys have only recently become
available, and one of them is the {\bf COMBO 17} survey ({\bf
C}lassifying {\bf O}bjects by {\bf M}edium-{\bf B}and {\bf
O}bservations in {\bf 17} filters, \citet{WolfCombo}), in some respect
the successor of CADIS. The complete catalogue will include $\sim
40000$ galaxies with $I\leq 23$, in 1\sqdegr, with $SED$ and
morphological information. This amazing data base can be used for
various investigations, using the projected correlation function. The
higher statistic allows for a more detailed analysis of the evolution
of the clustering of different galaxie types and their relation to the
underlying dark matter density field.

 The clustering properties of
starburst galaxies at higher redshifts can be investigated using the
emission line galaxies observed by the CADIS emission-line survey
using an imaging Fabry-Perot interferometer (Hippelein {\it et al.}
2002).These galaxies, which have been detected and classified by their
emission lines, have redshifts with an accuracy of 120\,km s$^{-1}$ --
good enough to calculate the three-dimensional correlation function
directly. The special observing technique samples galaxies in distinct
narrow redshift bins, which allows for the investigation of the
evolution of the clustering properties of emission line galaxies
between a redshift of $z\approx 1.4$ and $z\approx0.24$.

The observations have to be compared to theoretical predictions.
We plan to carry out a large cosmological simulation including
starformation and feedback. The individual galaxies will be assigned
an $SED$ according to their stellar masses and starformation
histories, so we can perform synthetic photometry applying our CADIS
or COMBO~17 filter set to the synthetic spectra. Thus we will be able
to ``observe'' the mock galaxies at different redshifts, and directly
compare the simulations to our observations.

\begin{acknowledgements}
We thank John~A. Peacock for many valuable and helpful discussions.\\
We thank all those involved in the Calar Alto Deep Imaging Survey,
especially H.-J. R{\"o}ser and C. Wolf, without whom carrying out the
whole project would have been impossible.\\
We are
greatly indebted to the anonymous referee who pointed out several points
which had not received sufficient attention in the original
manuscript. This led to a substantial improvement of the paper.\\
We also thank M. Alises and A. Aguirre for their help
and support during many nights at Calar Alto Observatory, and for carefully
carrying out observations in service mode.\\
\end{acknowledgements}

\bibliographystyle{aa}

\bibliography{lit}
\end{document}